\documentclass[11pt,a4paper]{article}
\pdfoutput=1 % if your are submitting a pdflatex (i.e. if you have images in pdf, png or jpg format)
\usepackage{jheppub}	% jheppub includes hyperref,color, natbib, amsmath, amssymb, epsfig, graphicx
 \usepackage{graphicx,amssymb,amsmath,amsfonts}
\usepackage{hyperref}
 \usepackage{diagbox}

  \usepackage{tikz}
\usetikzlibrary{positioning,arrows}
\usetikzlibrary{decorations.pathmorphing}
\usetikzlibrary{decorations.markings}
\usetikzlibrary{snakes}
\usetikzlibrary{matrix}
\usepackage{slashed}

\usepackage{graphicx}
 
\numberwithin{equation}{section}

\usepackage[utf8]{inputenc}

\date{\today}

\def\be{\begin{equation}}
\def\ee{\end{equation}}

\newmuskip\pFqmuskip

\newcommand*\pFq[6][8]{%
  \begingroup % only local assignments
  \pFqmuskip=#1mu\relax
  \mathchardef\normalcomma=\mathcode`,
  \mathcode`\,=\string"8000
  \begingroup\lccode`\~=`\,
  \lowercase{\endgroup\let~}\pFqcomma
  {}_{#2}F_{#3}{\left(\genfrac..{0pt}{}{#4}{#5}\Big | #6\right)}%
  \endgroup
}
\newcommand{\pFqcomma}{{\normalcomma}\mskip\pFqmuskip}
\newcommand{\pmat}{\begin{pmatrix}}
\newcommand{\fpmat}{\end{pmatrix}}
\newcommand{\eq}{\begin{equation}}
\newcommand{\feq}{\end{equation}}
\newcommand{\cas}{\begin{cases}}
\newcommand{\fcas}{\end{cases}}

\newcommand{\eqarray}{\begin{eqnarray}}
\newcommand{\feqarray}{\end{eqnarray}}

\newcommand{\f}{\phi}

\newcommand{\g}{\gamma}

\newcommand{\m}{\mu}
\newcommand{\n}{\nu}

\DeclareMathOperator\arctanh{arctanh}

				\def\g{\gamma}		\def\d{\delta}
						
		\def\k{\kappa}				\def\m{\mu}
\def\n{\nu}									\def\r{\rho}	
					\def\f{\phi}

						\def\L{\Lambda}

\def\be{\begin{equation}}
\def\ee{\end{equation}}
\def\bea{\begin{eqnarray}}
\def\eea{\end{eqnarray}}
\newcommand{\nn}{\nonumber}
\newcommand{\ft}[2]{{\textstyle\frac{#1}{#2}}}

\title{Gauge theory meets cosmology}

\author[a,b]{Massimo Bianchi}
\author[a,b]{, Giuseppe Dibitetto}
\author[b]{and Jose Francisco Morales}
\affiliation[a]{ Dipartimento di Fisica, Universit\`a di Roma ``Tor Vergata", Via della Ricerca
Scientifica 1, 00133, Roma, Italy.}
\affiliation[b]{INFN section of Rome ``Tor Vergata", Via della Ricerca
Scientifica 1, 00133, Roma, Italy.}

\emailAdd{bianchi@roma2.infn.it, dibitetto@roma2.infn.it, morales@roma2.infn.it}

\abstract{We reconsider linear perturbations around general Friedmann - Lemaitre - Robertson - Walker (FLRW) cosmological backgrounds. Exploiting gauge freedom involving only time reparametrizations, we write down classical background solutions analytically, for an arbitrary number of fluid components. 
We then show that the time evolution of scalar and tensor adiabatic perturbations are governed by Schr\"odinger-like differential equations of generalized Heun type. After recovering known analytic results for a single-component fluid, we discuss more general situations with two and three different fluid components, with special attention to the combination of radiation, matter and vacuum energy, which is supposed to describe the $\Lambda$CDM model. The evolution of linear perturbations of a 
 flat $\Lambda$CDM universe is described by a two-transient model, where the transitions from radiation to matter and matter to vacuum energy are governed by a Heun equation and a Hypergeometric equation, respectively.  We discuss an analytic approach to the study of the general case, involving generalized Heun equations, that makes use of (quantum) Seiberg-Witten curves for ${\cal N}=2$ supersymmetric gauge theories and has proven to be very effective in the analysis of Black-Hole, fuzzball and ECO perturbations. }

\begin{document}

\tikzset{
line/.style={thick, decorate, draw=black,}
 }

\maketitle

%\documentclass[11pt,a4paper]{article}
%%\usepackage{jheppub}	
%% jheppub includes hyperref,color, natbib, amsmath, amssymb, epsfig, graphicx
%
%\usepackage{amsmath,amssymb}
%\usepackage{latexsym}
%\usepackage{graphicx}
%\usepackage{epsfig}
%
%
%\def\ii{{\rm i}}
%
%\title{ Cosmological perturbations}
%
% 
%%\author[a]{Francesco Fucito,}
%%\author[a]{Jose Francisco Morales,}
%%%
%%%%
%%\affiliation[a]{Dipartimento di Fisica, Università di Roma ``Tor Vergata"  \& Sezione INFN Roma2, Via della ricerca
%%scientifica 1, 00133, Roma, Italy}
% 
%\abstract{  We study cosmological perturbations at linear order }
%
%\thispagestyle{empty} \clearpage
%
%\begin{document}
%\maketitle
%
%

\section{Introduction}
Since the turn of the millennium, a golden age for cosmology has started. This revolution was driven by a combination of interesting and compelling observations of both the early and the late universe.
One of the greatest achievements in the realm of observational cosmology nowadays is certainly the extraordinary precision with which the power spectrum of density
perturbations of the cosmic microwave background (CMB) has been measured over the past 25 years. 
On the other hand, around the same time, the surprising discovery of the accelerated expansion phase of the current universe \cite{SupernovaSearchTeam:1998fmf,SupernovaCosmologyProject:1998vns,Boomerang:2000jdg,Pryke:2001yz,SDSS:2003eyi,SDSS:2005xqv} has given rise to one the biggest challenges in theoretical physics: the origin of dark energy. 

The constant refinement of the observations of the large-scale structure of the universe, has promoted cosmology to a precision big-data driven science, at all scales. The standard model for present cosmology, the so-called $\L$CDM model, provides an accurate match to most observations. However, the increasing sensitivity of available measurements provide challenging stress-tests for this model, especially when observations carried out on the `early' universe are combined with those on the late universe. In particular a significant tension seems to take place between the values of the present Hubble constant $H_0$ measured  by Planck, based on the CMB, $H_0 \sim 68 \ \mathrm{km}\,\mathrm{s}^{-1}\mathrm{Mpc}^{-1}$ \cite{Planck:2018vyg} and by Dark Energy Spectroscopic Instrument (DESI) Collaboration \cite{Freedman:2021ahq,Freedman:2023jcz} or based on supernova surveys such as SH0ES (Supernova H0 for the Equation of State) \cite{Riess:2021jrx}, of about $H_0 \sim  74 \ \mathrm{km}\,\mathrm{s}^{-1}\mathrm{Mpc}^{-1}$. This `Hubble tension' suggests the need for new scenarios for dark matter, dark energy and other crucial ingredients in our understanding of the universe \cite{Abdalla:2022yfr,Kamionkowski:2022pkx,Verde:2023lmm,Lynch:2024hzh}. 

Another cosmological observable which is currently subject to an observational tension is the so-called $\sigma_8$ parameter. This quantity describes the galaxy distribution in the celestial sky and it is related to the linear matter density fluctuations $\delta\rho_{\text{m}}/\rho_{\text{m}}$  within a distance scale of $8 h^{-1}$ Mpc, with $h=H_0/100$ Km/s/Mpc. The evolution in time of this quantity
is sensitive to the details of the assumed cosmological model.
Since generic completions of $\Lambda$CDM that alleviate the Hubble tension usually tend to worsen the $\sigma_8$ tension, it may become extremely important to design high-precision tests for different models, based on their predictions for cosmological perturbations \cite{Abdalla:2022yfr,Kamionkowski:2022pkx,Verde:2023lmm,Lynch:2024hzh}.
 %We will not address these hot issues that have been discussed at length and in details recently\footnote{See e.g. the talks on this topic at  the Lemaitre Conference in June, available in the website.}.
 
  The main aim of our present investigation is giving a new perspective and new techniques to approach the study of cosmological perturbations.  We consider a classical Friedmann-Lemaitre-Robertson-Walker (FLRW) 
expanding universe coupled to a general multi-component perfect fluid. Exploiting time reparametrization freedom, we describe the cosmological evolution by using the size $a$\footnote{
The red shift z  is related to the scale factor $a$  by z+1=$1/a$.}
 of the three-dimensional universe slice  as our time coordinate\footnote{We thank Misao Sasaki for the suggestion during the Lemaitre conference in June at Specola Vaticana.}. 
With this choice the classical background solution can be written in analytic form, for an arbitrary number of fluid components. In the specific case of the $\L$CDM model, we find an explicit expression for the deceleration function $q(a)$ that describes a universe comprising two transients dividing the three main era dominated by radiation, matter and vacuum energy, respectively. We study the time-evolution of scalar and tensor linear perturbations. We find that adiabatic perturbations are always described by Schr\"odinger-like differential equations with a number of Fuchsian singularities depending on the number and type of fluid components. Linear perturbations for single component universes are known to be described by Bessel functions \cite{Mukhanov:1990me,Vittorio:2017foh,Kodama:1984ziu}. For two component universes we find Hypergeometric and Heun equations, for three and four components we find equation with five to seven singularities, that we will refer to as {\it generalized Heun equations}. Analytic solutions for adiabatic perturbations of the two-transient $\L$CDM model are explicitly written by gluing local solutions along the two transients. 

(Confluent) Heun equations are known to describe the dynamics of linear perturbations around black holes \cite{Regge:1957td,Teukolsky:1972my,Mino:1997bx,Sasaki:2003xr}.
The solutions and connection formulae for Heun equation and its confluences have been recently derived in \cite{Aminov:2020yma, Bianchi:2021xpr,Bonelli:2021uvf,Bianchi:2021mft,Bianchi:2021yqs,Bonelli:2022ten} based on CFT and gauge theory inspired techniques \cite{Belavin:1984vu, Nekrasov:2002qd,Flume:2002az,Bruzzo:2002xf,Alday:2009aq,Nekrasov:2009rc, 
Poghossian:2010pn,Fucito:2011pn}. Generalized Heun equations have been put in correspondence with quantum Seiberg Witten (SW) curves and solutions related to partition functions of linear quiver gauge theories  \cite{Consoli:2022eey}. 
The techniques have been applied to the study of binary systems, fuzzballs and ECOs  \cite{Bianchi:2022wku,Bianchi:2022qph,Bianchi:2023rlt,Bianchi:2023sfs,Fucito:2023afe, DiRusso:2024hmd,Cipriani:2024ygw,Bianchi:2024vmi,Bena:2024hoh}.  Here we apply the SW gravity correspondence to the study of  generalized Heun equations  describing the evolution of adiabatic perturbations for universes filled in with multi-component fluids. In particular, we write the explicit dictionary for  scalar and tensor perturbations of a two-component universe made of radiation and matter.

  The paper is organised as follows: in section 2 we study background solutions for FLRW universes with arbitrary number of fluid components. In section 3, we study the evolution of adiabatic perturbations at linear order. In section 4 we describe the evolution of cosmological perturbations of the $\L$CDM two-transient model. In section 5 we introduce the quantum Seiberg-Witten / Cosmology correspondence, which is a new form of gauge / gravity duality, similar to the one between BH and fuzzball perturbations and ${\cal N}=2$ linear quiver theories, yet different from  the holographic cosmology originally proposed in \cite{Skenderis:2006jq,Skenderis:2007sm} and further elaborated on in \cite{McFadden:2009fg,Bzowski:2023nef}, and work out explicitly the dictionary for the case of a two-component universe made of radiation and matter. In Section 6 we draw some conclusions and comment on open problems and future directions.

\section{The dynamics of an FLRW universe revisited}
\label{section:FLRWbackrgounds}
Let us consider the general (Friedmann-Lemaitre-Robertson-Walker) FLRW metric describing a homogenous and isotropic universe\footnote{Alternatively, one can write
\be
\label{FLRW_0}
ds_4^2 \,=\,a(\eta)^2( -d\eta^2 +  \, ds_{\mathcal{M}_3}^2) \qquad , \qquad d\eta={dt\over a(t)}
\ee
 with $\eta$ the conformal time. 
}
\be
\label{FLRW_0}
ds_4^2 \,=\, -dt^2 + a(t)^2 \, ds_{\mathcal{M}_3}^2 \ ,  
\ee
where $t$ is cosmic time, $a(t)$ is the scale factor describing how spatial lengths evolve throughout cosmic history, and the spatial 3D slices $\mathcal{M}_3$ are maximally symmetric, \emph{i.e.} they have constant curvature. Thus, depending on whether this is zero, positive or negative, respectively we have $\mathbb{R}^3$, $S^3$ or $\mathbb{H}^3$. 
In  cartesian coordinates, the spatial metric may be expressed as
\be
ds_{\mathcal{M}_3}^2 \,=\,  {d x_i^2 \over 1+\ft{ \kappa M_{\text{Pl}}^2}{4}  x_i^2} \ ,
\label{ds2_3D}
\ee
where the scale $M_{\text{Pl}}$ has been inserted in order to get a \emph{dimensionless} discrete curvature parameter $\kappa=+1,0,-1$. We work in natural units where $\hbar = c=1$, hence $M_{\text{Pl}}^2 = (8\pi G_{\text{N}})^{-1}$.

A perfect fluid coupled to the above background metric generates a stress-energy tensor of the form
\be
\label{T_fluid}
 T_{{\mu}{\nu}}=(\rho+p)\, u_{\mu} u_{\nu} +p\, g_{{\mu}{\nu}}  
% T_{00}=a(\eta)^2\left[ 1+2 \Phi(\eta,r) \right] \rho(\eta) \qquad , \qquad   T_{ii}=  a(\eta)^2\left[ 1-2 \Phi(\eta,r) \right] p(\eta)
 \ee
 where $p$ is the pressure, $\rho$ the energy density and $u^{\mu}$ the quadrivelocity of the fluid. The above stress-energy tensor assumes the diagonal form $\mathrm{diag}(\rho,p,p,p)$ when expressed in its locally inertial frame (LIF), where the metric is flat and the quadrivelocity is $u^\mu=(1,0,0,0)$. If one now specifies an equation of state of the form \be 
 p \,=\, w\rho
 \ee
the corresponding scale factor of a universe coupled to this fluid has a power-law behavior in cosmic time, \emph{i.e.} 
\be 
a(t) \sim t^{\frac{2}{3(1+w)}}
\ee
with $w>-1$. The extreme case $w=-1$ represents vacuum energy and is solved by a scale factor that grows exponentially fast in $t$. If one moves to more involved settings with more than one fluid, then the Einstein equations can no longer be integrated analytically and one usually proceeds with piecewise solutions describing different epochs in cosmic history, each one characterized by the domination of a single fluid component.
In the next paragraph we will exploit local time reparametrization freedom in order to use the scale factor $a$ as time variable and obtain global analytic solutions for backgrounds coupled to multi-component fluids.

\subsection{A useful time coordinate}

As just anticipated, we use the size of the three-dimensional slice $a$ as time coordinate   and write the metric \eqref{FLRW_0}   as
\be
\label{bckd}
ds_4^2 \,=\,  -\frac{da^2}{a^2 b(a)^2} + a^2 \,ds_{\mathcal{M}_3}^2 \\ ,
\ee
with $a$ related to the cosmological time via 
\be
dt ={d  a  \over a b(a) }  
\ee
In an expanding universe,  $a$ grows from $0$ in the far past, to $a=1$ today, and $a=\infty$ in the far future.  
On the other hand 
 \be 
 b(a)={\dot{a}\over a} =H(a) \, ,
 \ee
encoding the  {\it Hubble rate} as a function of the scale factor $a$, is determined by   Einstein equations
 \be
 G_{\m\n} \,=\, R_{{\mu}{\nu}}-\ft12 g_{{\mu}{\nu}} R \,=\, M_{\text{Pl}}^{-2} T_{{\mu}{\nu}} \ ,
 \ee
 for a given stress energy tensor. We consider a universe filled in with a perfect fluid, so
that the  stress-energy tensor  is given by \eqref{T_fluid}  
 with quadrivelocity\footnote{$u^a = u^t \dot{a} = \dot{a} = a b(a) = a H(a)$, thus $u_a = g_{aa}u^a = - (ab)^{-1}= -(aH)^{-1}$.} 
\be
 u_{\mu} = g_{{\mu}{\nu}} u^{\nu} = \left(-\ft{ 1}{a\, b(a)} ,0,0,0\right) \ ,
\ee
Assuming that $p(a)$ and $\rho(a)$ depend only on $a$ and denoting by $\prime$ (prime) derivatives w.r.t. $a$, the Einstein equations can be written as
\bea
 p(a) \,&=& \, -\rho(a)-{a\, \rho'(a) \over 3 } \ , \nn\\
 b(a) &=&  \sqrt{  \frac{\rho(a) }{3 M_{\text{Pl}}^2 }   - {M_{\text{Pl}}^2 \kappa\over  a^2}} \label{rhobEE}
\eea
in terms of the energy density $\rho(a)$.

 \subsection{What's the (cosmic) time?}

 As we mentioned, cosmic time $t$ tends to complicate the analysis of both the background solution and the linear perturbations. Yet we should check that $a(t)$ be monotonous if we want it to be a `good' time variable. The relation between $t$ and $a$ is determined by 
 \be
 dt = {da a\over b(a)} = {da \over a H(a)} =  {da \over a H_0 \sqrt{\sum\limits_{i}   \Omega_i  a^{-3(1+w_i)}  +\Omega_\kappa a^{-2}}}
 \ee
that in general cannot be expressed in terms of elementary functions except for single and some two- and three-component fluids, discussed in appendix \ref{Cosmic_t}.

In the $\Lambda$CDM model, one has
\be
H_0 t = \int {da \over \sqrt{\Omega_\Lambda a^2 +\Omega_\kappa + \Omega_{\rm m} a^{-1} + \Omega_\gamma a^{-2}}} =  \int {a da \over \sqrt{\Omega_\Lambda a^4 +\Omega_\kappa a^2 + \Omega_{\rm m} a + \Omega_\gamma}}
\ee
that can be written in terms of elliptic functions. 

\begin{figure}[h!]
  \begin{minipage}{\textwidth}
    \centering
\includegraphics[width=0.48\textwidth]{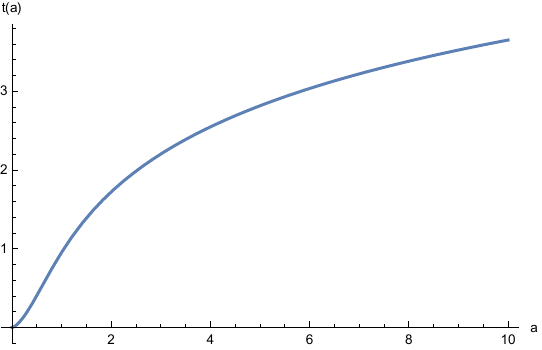}
    \caption{\it  Plot of cosmic time $t$ as a function of scale factor $a$ in the $\Lambda$CDM model with $\Omega_\Lambda  =0.6889$, $\Omega_{\text{m}}  =0.3111$, $\Omega_\gamma =4.6350 \times 10^{-5}$, $\Omega_\kappa =0$.}
    \label{timevsaplot}
  \end{minipage}
\end{figure}

In Fig.\ref{timevsaplot} we plot cosmic time $t$ as a function of $a$ for the phenomenologically viable choice of fluid abundances
\be
\label{PhenoOmega}
 \Omega_\Lambda  =0.6889 \qquad , \qquad   \Omega_{\text{m}}  =0.3111 \qquad , \qquad   \Omega_\gamma =4.6350 \times 10^{-5} \ ,\nn
 \ee
with $\Omega_\kappa =0$.
 
\subsection{A single component universe}
Let us start by considering a universe in the presence of a single-component fluid. It may be worth mentioning that this ideal situation is also suited for describing universes with a richer structure, in the limit where their dynamics is dominated by a single fluid in flat space, {\it i.e.} $\kappa=0$. In this case, the energy density scales as $\rho\sim a^{-n}$ and (\ref{rhobEE}) yields 
\be
\rho \,=\,3 (M_{\text{Pl}}\, H_0)^2  \, a^{-n} \quad  , \quad  p \,=\, 3 w\, (M_{\text{Pl}}\, H_0)^2   \, a^{-n} \quad , \quad 
b^2 \,=\,  H_0^2  \, a^{-n} \quad, \quad n=3(1+w)   \label{p1comp}
\ee
with $H_0=b(1)$ the Hubble constant today. 
 The general feature of universes coupled to a single fluid is that of having a constant speed of sound 
 \be
 c_s^2(a) \,=\, \left({ \partial p\over \partial \rho}\right)_s=w
 \ee
  The results for some relevant choices of $w$ are given in table \ref{Table:w_fluids}. It is worth mentioning that the contribution of the spatial curvature of the 3D slices $\kappa$, once taken to the RHS of Einstein equations, can be re-interpreted as the energy density of a perfect fluid with  $w=-\frac{1}{3}$ and 
  \be
   \rho_\kappa  =-\,3  \kappa (M_{\text{Pl}}^2\, H_0)^2 \, a^{-2}  
   \ee
    The same equation of state describes a gas of freely propagating strings, that will be labelled by the letter $\sigma$. One could set $\kappa=0$ and reabsorb the contribution of the curvature inside $\rho_\sigma$ (with `string tension' $\sigma \sim -\kappa M_{\text{Pl}}^2$) but we will refrain from doing that.

   \begin{table}[t]
  \label{tabcomp}
\begin{center}
\begin{tabular}{|c|c|c|c|c|c|}
\hline
 \text{Fluid Type} & Symbol & $ w$ & $n$ &   $b(a)$ &$ \eta(a) $ \\
 \hline\hline
 \text{Vacuum} & $\L$ & $-1$ & $0$   & $1$  &$ -1/a $  \\
\hline
\text{Strings, Curvature} & $\sigma,\kappa$ & $-\frac{1}{3}$ & 2  & $ a^{-1}   $&$  \log(a) $ \\
\hline
 \text{Matter} & m & $0$ & $3$   &$ a^{-3/2} $&$ 2 a^{{1/2}}$ \\
\hline
 \text{Radiation} & $\g$ & $\frac{1}{3}$ & $4$   &$ a^{-2} $&$ a$ \\
 \hline
\text{Stiff}  & $s$ & 1 & 6 &  $a ^{-3}$ & $ a^2/2$ \\
 \hline
\end{tabular}
\end{center}
\caption{Cosmologies for universe made of a single-component perfect fluid  in flat space. We set $\Omega H_0=1$ for simplicity.}
\label{Table:w_fluids}
\end{table}% 

 \subsection{A multi-component universe}
If a universe is instead coupled to multiple and non-interacting components of different kinds, we make the following \emph{Ansatz} for the stress-energy tensor
\be
\label{Tmunu_0}
%\left\{
\begin{array}{lclc}
\rho(a) & = & 3 (M_{\text{Pl}}\, H_0)^2  \sum\limits_{i} \Omega_i a^{-3(1+w_i)}  & , \\
p(a) & = &3 (M_{\text{Pl}}\, H_0)^2\,  \sum\limits_{i} w_i\,\Omega_i a^{-3(1+w_i)}  & , 
\end{array}
%\right.
\ee
where $ i=\Lambda, \sigma,m,\gamma,s$ runs over the fluid components, $w_i$ is the equation of state parameter of the $i$-th species and $\Omega_i$ measures the abundance of the $i$-component at $a=1$, \emph{i.e.} today in our conventions.  
 The general solution for the {\it Hubble rate} $b(a)=H(a)$ then reads
\be
\label{gmunu_0}
b^2(a) \ = \ H_0^2  \left( \sum\limits_{i}   \Omega_i  a^{-3(1+w_i)}  +\Omega_\kappa a^{-2}\right)  
\ee
with $i$ running over the fluid components as in \eqref{Tmunu_0} and $\sum_i \Omega_i=1-\Omega_\kappa$.

\begin{figure}[h!]
  \begin{minipage}{\textwidth}
    \centering  
\includegraphics[width=0.48\textwidth]{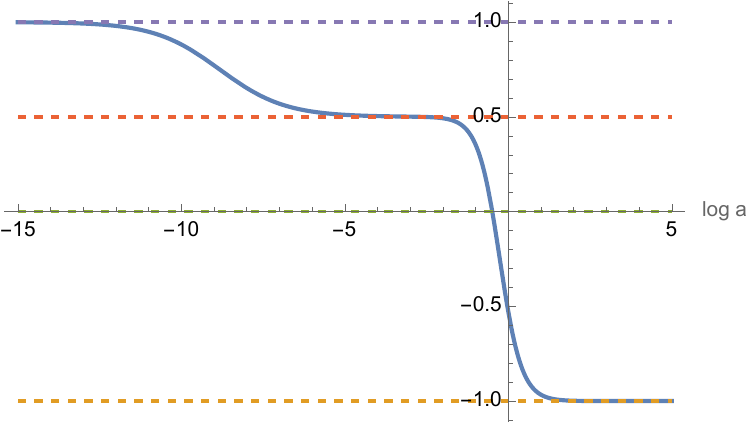}
    \caption{\it  Decelaration parameter $q$ for $\kappa=0$ in the $\Lambda$CDM model with $\Omega_\Lambda  =0.6889$, $\Omega_{\text{m}}  =0.3111$, $\Omega_\gamma =4.6350 \times 10^{-5}$. Time evolution is plotted w.r.t. $log(a)$. Remote past is far left, while far future is extreme right.}
    \label{figadec}
  \end{minipage}
\end{figure}

For our current description of the universe in terms of the $\L$CDM model, we need three different fluids: vacuum energy $\Lambda$ ($w_\L=-1$), matter m ($w_{\text{m}}=0$), radiation $\gamma$ ($w_\g=1/3$),  since the curvature contribution ($w_\kappa=-\ft13$) is compatible with $\kappa=0=\Omega_\kappa$ and the contribution of other fluids is negligible as well. The corresponding expressions for the energy density and pressure take the form
 \be
%\left\{
\begin{array}{lclc}
 \rho(a) &=&3(M_{\text{Pl}}\, H_0)^2 \, \left(  \Omega_\Lambda + \Omega_{\text{m}} a^{-3} +\Omega_\gamma a^{-4}  \right) & ,\\[2mm]
   p (a)&=& 3(M_{\text{Pl}}\, H_0)^2 \, \left(-\Omega_\Lambda +\ft13 \Omega_\gamma a ^{-4}  \right) & .
\end{array} 
%\right.
\label{rhotau} 
 \ee
 The solution for the $\L$CDM {\it Hubble rate} reads
\be
b^2 \ = \  H_0^2 \left(  \Omega_\g \,a^{-4}+\Omega_{\text{m}}\,a^{-3}+\Omega_\L +\Omega_\kappa\, a^{-2}\right)
\ee
An efficient way of characterizing an epoch during cosmic history relies on the so-called deceleration parameter
 \be
 q(a)\,\equiv\,-{\ddot{a} a \over \dot{a}^2} \,=\, -1-\frac{ a\, \partial_a b(a)^2}{2 b(a)^2 }\ ,
 \ee
For general FLRW backgrounds, with our choice of time coordinate, the deceleration parameter explicitly evaluates to 
\be
 q(a)\,=\,\frac{\Omega_\gamma a^{-4}+\ft12 \Omega_{\text{m}} a^{-3} -\Omega_\Lambda} {\Omega_\gamma a^{-4}+ \Omega_{\text{m}} a^{-3} -\Omega_\k a^{-2} +\Omega_\Lambda }\ .
\ee
 In Fig.~\ref{figadec}, we display the deceleration parameter for the phenomenologically interesting choices \cite{Planck:2018vyg}  
in the spatially flat case $\k=0$.
 From Fig.~\ref{figadec},  one notices that the history of universe can be mainly divided into two transients: the first one interpolating between a radiation dominated era ($ a\ll 1$) to a matter dominated one ($  -5 \lesssim \log a  \lesssim -2$), and a second one towards a future universe dominated by the cosmological constant $ a\gg 1$.  

 \section{Linearized perturbations around an FLRW background}
Now that the stage is set, we are ready for discussing cosmological perturbations at a linearized level. Due to the local invariance of General Relativity under diffeomorphisms, one first needs to pick out a physical set of metric perturbations after removing gauge redundancies and combine them into irreducible pieces. Our perturbed metric in general reads
\be
g_{\m\n} \,=\, \underbrace{{g}^{(0)}_{\m\n}}_{\text{hom. } + \text{ iso.}} \,+\, \ \delta g_{\m\n} \ ,
\ee
where ${g}^{(0)}$ is a homogeneous and isotropic background of the class discussed in the previous section, while $\d g$ is a small perturbation. Such perturbations decompose into scalar, vector and tensor modes.

The general form of scalar perturbations may be written as
\be
\d g_{\m\n}^{\text{(S)}} \,=\, a^2\,\left(\begin{array}{c|c}-2 a^{-4} b^{-2}\f & a^{-2}b^{-1}\nabla_jB \\ \hline
 a^{-2} b^{-1}\nabla_iB & -2\left(\psi\g_{ij}-\nabla_i\nabla_jE\right)\end{array}\right) \ ,
\ee
where $\phi$, $\psi$, $B$, $E$ are scalar functions and all derivatives are assumed to be covariant w.r.t. ${g}^{(0)}$. 

Vector perturbations may be parametrized as 
\be
\d g_{\m\n}^{\text{(V)}} \,=\, a^2\,\left(\begin{array}{c|c}0 & a^{-2} b^{-1}S_j \\ \hline
 a^{-2} b^{-1}S_i & -2\nabla_{(i}F_{j)}\end{array}\right) \ ,
\ee
where both $F$ and $S$ are divergence-free 3D vector fields. 

Finally, tensor perturbations are of the form
\be
\d g_{\m\n}^{\text{(T)}} \,=\, a^2\,\left(\begin{array}{c|c}0 & 0\\ \hline
0 & h_{ij}\end{array}\right) \ ,
\ee
where $h$ is transverse and trace-less {\it i.e.} satisfies $h_i{}^i \,=\, \nabla^jh_{ij} \,=\, 0$.

As far as purely scalar perturbations are concerned, there are two independent diffeomorphisms that preserve the scalar nature of these perturbations without inducing any mixing with vector and tensor modes. This implies that only two physical combinations of $(\f,\psi, E, B)$ survive after removing gauge redundancies. These are usually denoted by $(\Phi,\Psi)$ and they generalize $(\f,\psi)$ outside of the so-called \emph{longitudinal gauge}, where $E=B=0$. Adopting the longitudinal gauge from now on, the scalar-perturbed metric may be cast into the form
\be
ds_4^2|_\textrm{S} \, \overset{\textrm{L}}{=}\,  -\left(1+2 \Phi(a,\bold{x}) \right)\frac{da^2}{a^2 \, b(a)^2} + a^2 \left(1-2 \Psi(a,\bold{x}) \right)ds_{\mathcal{M}_3}^2 \ ,
\ee
where $\bold{x}$ are the coordinates on the 3D spatial slices of our universe. It may be worth mentioning that a further restriction to isotropic sources for these perturbations, \emph{i.e.} $\d T^i{}_j \propto \d^i_j$, leads to perturbations that only depend on $a$ (time) and $r$ (radial coordinate), together with the extra constraint $\Psi=\Phi$. In this particular setup $\Phi$ may be interpreted as a generalized Newtonian potential.

On the other hand, the tensor-perturbed metric may be written as
\be
ds_4^2|_\textrm{T} \, =-\frac{da^2}{ a^2 \, b(a)^2} +\,  a^2 \,\left( ds_{\mathcal{M}_3}^2+h_{ij}(a,\bold{x})\,dx^idx^j \right)\ ,
\ee
with $h$ still subject to  $h_i{}^i \,=\, \nabla^jh_{ij} \,=\, 0$.

\subsection{Scalar gauge-invariant perturbations}
\label{subsection:scalar_pert}
Let us now focus on isotropic scalar perturbations around FLRW backgrounds of the form in \eqref{bckd}. As already mentioned above, choosing the (generalized) longitudinal gauge further specifies the perturbed metric to be of the form 
\be
ds_4^2|_\textrm{S} \, \overset{\textrm{L}}{=}-\left(1+2 \Phi(a,\bold{x}) \right)\frac{da^2}{a^2 \, b(a)^2} + a^2 \left(1-2 \Psi(a,\bold{x}) \right)ds_{\mathcal{M}_3}^2 \ ,
\ee
with $ds_{\mathcal{M}_3}^2$ given in \eqref{ds2_3D}. We take
\bea
u_\mu( a,{\bf x})  dx^\mu & =&  \frac{ \sqrt{1+2 \Phi(a, {\bf x} )}  }{ a\, b(a)}\, da \,+\, \d u_i dx^i  = {u}_0 + \delta{u}\ , \nn\\
p(a)  &=&  p_0(a) \,+\, \delta p(a) \nn\\
\rho(a)  &=& \rho_0(a) \,+\, \delta \rho(a) 
\eea
with
\bea
 p_0(a)  \,&=& \, -\rho_0(a)-{a\, \rho_0'(a) \over 3 } \qquad , \qquad 
 b(a) =  \sqrt{  \frac{\rho_0(a) }{3 M_{\text{Pl}}^2 }  - {M_{\text{Pl}}^2 \kappa \over a^{2}} } \label{rhob}
\eea
 Plugging this \emph{Ansatz} into the Einstein equations and expanding up to linear order in $\Phi$, $\d\r$, $\d p$ and $\delta u_i$,  one finds the following system of second order partial differential equations (PDE)
 \bea\label{PDEforPhi}
 && 
  \partial_a \partial_i \left(a\,\Phi\right) \,=\, \frac{ 2a(\r_0+p_{0}) \d u_i }{ b M_{\textrm{Pl}}^2 (4 + M_{\textrm{Pl}}^2 \kappa x_i^2) } \ ,\nn\\[2mm]
 &&3 a^3 b^2 \, \partial_a\Phi -\Delta\Phi+\left(3 a^2 b^2  -3 \kappa M_{\textrm{Pl}}^2 \right)\Phi+\frac{a^2 \delta \rho}{2  M_{\textrm{Pl}}^2}\,=\,0 \ ,\\
   && {b\over a}  \,\partial_a ( a^5 b  \partial_a \Phi) +  \left[  \partial_a (a^3 b^2)-  \kappa M_{\textrm{Pl}}^2  \right] \Phi -\frac{a^2 {\delta p}}{2 M_{\textrm{Pl}}^2}\,=\,0\ ,\nn
 \eea
The first equation  determines $\d u_r$ once the Newtonian potential $\Phi$  has been determined.
% with\footnote{Where $k^2=|\vec{k}|^2$ for $\kappa=0$, $k^2=\ell(\ell+1)$ for $\kappa=+1$ and $k^2=h(h-1)?$  for $\kappa=-1$.} 
We study the time evolution of each Fourier component of $\Phi$, for which\footnote{We use plane-wave notation but the ansatz can be used for $\kappa\neq 0$, replacing $-k^2$ with the eigenvalue of the relevant  laplacian $\Delta_{{g}^{(0)}}$.}
  \be  
\Phi(a,{\bf x})=e^{{\rm i} {\bf k} {\bf x} } \Phi(a)  \qquad \Rightarrow \qquad \Delta \Phi= - k^2 \Phi \ ,
 \ee

 We further focus on \emph{adiabatic perturbations} and neglect entropy or iso-curvature modes. It is known\footnote{We thank Misao Saski for reminding us about this issue} \cite{Kodama:1984ziu, Mukhanov:1990me, Vittorio:2017foh} that adiabatic modes are necessarily coupled to entropy (or isocurvature) modes in the case of a multi-component fluid. In particular the coupling with these modes introduces non-homogeneous source terms in the equations for the adiabatic modes that are particularly relevant during the transient from one fluid phase to another. Discarding these effects for the time being, the variations of  pressure and energy density are related  via  
 \be
 \delta p(a) =c_s^2(a) \delta \rho(a) 
 \ee
 with
 \be
  c_s^2(a)={p'(a)\over \rho'(a)} =-\frac{4}{3} -{a\rho_0''(a) \over 3 \rho_0'(a) } 
 \ee
 For this choice the last two equations combine into the following homogeneous second order ordinary differential equation (ODE)
\bea
\frac{b}{a}  \partial_a\left(a^5 b\, \Phi '\right) +3 a^3 b^2 c_s^2 \Phi'+ \left[ \partial_a \left(a^3 b^2\right)-\kappa  M_{\textrm{Pl}}^2+c_s^2 \left(k^2+3 a^2 b^2-3 \kappa  M_{\textrm{Pl}}^2\right)\right] \Phi  =0
%&&2 a^2 \rho _0' \Phi '' \left[a^2 \rho _0-3 \kappa  M_{\text{Pl}}^4\right]+\Phi ' \left[a^4 \left(\rho _0'\right){}^2+2 a^3 \rho _0 \rho
%   _0'+2 a^2 \rho _0'' \left(3 \kappa  M_{\text{Pl}}^4-a^2 \rho _0\right)\right] \\
% &&  +\Phi  \left[2 a^3 \left(\rho _0'\right){}^2-2 \rho _0' \left(a^2 \rho _0+4 k^2 M_{\text{Pl}}^2-18 \kappa 
%   M_{\text{Pl}}^4\right)-2 a \rho _0'' \left(a^2 \rho _0+k^2 M_{\text{Pl}}^2-6 \kappa  M_{\text{Pl}}^4\right)\right]=0\nn
 \label{eqphi}
 \eea
where primes denote derivatives wrt  $a$, as always. 
The equation can be written in canonical Schr\"odinger-like form\footnote{Even though $a$ is a time-like variable.}
 \be
 \Psi''(a) + Q(a) \Psi(a)=0 \label{eqcan}
 \ee
 with
 \be
 \Phi(a)={ \sqrt{ \rho_0'(a)} \over ( a^2 \rho_0(a)-3 \kappa M_{\text{Pl}}^4 )^{1\over 4} }   \Psi(a)
 \ee
 and
 \bea
% Q(a) &=&+\frac{\rho ^{'''}}{2 \rho '}-\frac{3 \left(\rho ''\right)^2}{4 \left(\rho '\right)^2} + \frac{27 a^4 \left(\rho '\right)^2}{16 \left(\kappa -3 a^2 \rho \right)^2}+\frac{27 a^3 \rho  \rho '}{4 \left(\kappa -3 a^2 \rho \right)^2}+\frac{\rho '' \left(3 a^2 \rho -4 \kappa +2
%   k^2\right)}{\rho ' \left(2 a \kappa -6 a^3 \rho \right)} \nn\\
%   && +\frac{-27 a^4 \rho ^2+6 a^2 \rho  \left(15 \kappa -8 k^2\right)+8 \kappa  \left(2 k^2-3 \kappa \right)}{4 \left(a \kappa -3
%   a^3 \rho \right)^2}
Q(a) &=& \frac{\rho _0'''}{2 \rho _0'}-\frac{3 \left(\rho _0''\right){}^2}{4 \left(\rho
   _0'\right){}^2}   -\frac{k^2 M_{\text{Pl}}^2 \left(a \rho _0''+4 \rho _0'\right)}{a^2 \rho _0' \left(a^2 \rho _0-3 \kappa  M_{\text{Pl}}^4\right)}
   +\frac{\rho _0'' \left(12 \kappa  M_{\text{Pl}}^4-a^2 \rho _0\right)}{2 a \rho _0' \left(a^2 \rho _0-3 \kappa  M_{\text{Pl}}^4\right)} \nn\\
   && +\frac{3 \left(a^6 \left(\rho _0'\right){}^2+4 a^5 \rho _0 \rho _0'-4 a^4 \rho _0^2+120 a^2 \kappa 
   \rho _0 M_{\text{Pl}}^4-288 \kappa ^2 M_{\text{Pl}}^8\right)}{16 a^2 \left(a^2 \rho _0-3 \kappa  M_{\text{Pl}}^4\right){}^2} \label{qq0}
   \eea
   In the rest of this subsection, we will discuss some analytic solutions of this ODE for various choices of fluid components.
\subsubsection*{A single component universe}
 In the simple case of a single component universe, 
 \be
 \rho_0(a)=3 (M_{\text{Pl}}\, H_0)^2 \, a^{-3(1+w) }  
 \ee
  the linearized field equation for the scalar adiabatic perturbation (\ref{eqphi}) becomes
 \be
 \label{lineq1}
 a^{1-3 w} \Phi ''(a)+\frac{1}{2} (3 w+7) a^{-3 w} \Phi '(a)+\ft{k^2 w}{H_0^2}   \Phi (a)=0 \ ,
 \ee
 The general solution of (\ref{lineq1}) for $w\neq 0,-\ft13$ is given by\footnote{ The asymptotics follows from those of the Bessel functions
 \be
 J_\alpha (x) \underset{x \to 0}{\approx}  {x^\alpha \over 2^\alpha \Gamma(1+\alpha) } +\ldots   \qquad , \qquad   J_\alpha (x) \underset{x\to \infty}{\approx}  \sqrt{ \ft{2}{\pi x} } \cos(x-\ft{\pi}{4}-\ft{\alpha}{2} 
 )+\ldots  
 \ee  }
 \be
\label{sol_1CompPhi}
\hspace{-3mm}\Phi (a)\,=\, a^{-\frac{5+3w}{4}} \left[c_1 \, J_{\alpha}\left(\gamma a^{1+3w\over 2} \right)+c_2 \, J_{-\alpha}\left( \gamma a^{1+3w\over 2} \right)\right]
 \ee
 with $c_1$ and $c_2$ arbitrary real constants,
 \be
 \alpha=\frac{5+3w}{2 (1+3w)} \ , \qquad  , \qquad \gamma={ 2   k \sqrt{w} \over  H_0 (1+3 w)} 
 \ee
and $J_{\pm\alpha}$  the (modified) Bessel function.  For $w=0,-\ft13$ one finds instead
\bea
w&=&0:  ~~~\qquad\qquad \Phi = c_1+c_2 a^{-{5\over 2}} \nn\\
w&=&-\ft13: \qquad\qquad \Phi  = c_1 a^{-1+\sqrt{1+ {k\over 3\Omega} }} +c_2 a^{-1-\sqrt{1+ {k\over 3\Omega} }}   
\eea
%\footnote{These Bessel functions become the modified ones if the equation of state parameter $w$ gets negative. It may also be worth noticing that the $w=0$ and $w=-\frac{1}{3}$ cases need to be studied separately}.  
 We notice that $\alpha>0$ for $w>-1/3$ and 
  $J_{-\alpha}$ is singular for $ a\to 0$, which implies $c_2=0$ if one is interested in smooth initial conditions in the far past\footnote{For $-1/3>w>-1>-5/3$ instead, $\alpha<0$ and   $J_{+\alpha}$ is singular for $ a\to 0$.}. The solution \eqref{sol_1CompPhi} exactly reproduces the one in \cite{Mukhanov:1990me,Vittorio:2017foh,Kodama:1984ziu} given in terms of the conformal time $\eta$, related to cosmic time $t$ and $a$ via
 \be
 \label{tandetaoftau}
d\eta=  {dt\over a(t)} = {da  \over a^2 b(a)}  = da \sqrt{  \frac{3 M_{\text{Pl}}^2 }{ a^4 \rho(a) -3M_{\text{Pl}}^4\kappa a^{2}} }
 \ee 
 that for a single component fluid yields
 \be
  \eta = 
\left\{
\begin{array}{ccc}
H_0^{-1} {2 \, a^{\frac{1+3w}{2}}\over  (1+3w)}   &~~~~   &  w\neq -\ft13 \\
H_0^{-1}  \log a& ~~~~   &  w=-\ft13    \\  
\end{array}
\right.
  \ee

Long wavelength perturbations, \emph{i.e.} $c_s^2\,k^2 \gg 3 M_{\text{Pl}}^2 $ remain stable after entering the Hubble horizon, while short wavelength perturbations become sound waves beyond the Hubble horizon and the associated Newtonian potential $\Phi$ undergoes a phase of damped oscillations with an amplitude that decays as a power-law in $a$ towards far future. This phenomenon is also known as the Jeans instability that can trigger structure formation \cite{Mukhanov:1990me,Vittorio:2017foh}. A similar transition when moving from long to short wavelengths also turns out to occur in cosmological backgrounds with multi-component fluids, as we will explore next.
 
 \subsubsection*{A multi-component universe}
 
 Linear perturbations around a multi-component  universe are described by the Schr\"odinger-like equation (\ref{eqcan}) with singularities at $a=0,\infty$ and at the solutions of
\be
  \rho_0'(a) \left(a^2 \rho_0(a)-3 \kappa  M_{\text{Pl}}^4\right)=0
\ee
  The number of singularities depends on the specific choice of $\rho_0(a)$. For example, the various phases of the $\L$CDM cosmology model involve three different fluids (radiation, matter, vacuum) with energy density
\be 
\rho_0(a)  =  3 (M_{\text{Pl}}\, H_0)^2\, \left(  \Omega_\Lambda + \Omega_{\text{m}} a^{-3} +\Omega_\gamma a^{-4}  \right)  
\ee
 The pattern of singularities for the various choices of multi-component universes, including also the presence or absence of a non-trivial 3D curvature $\kappa$ (or equivalently a gas of strings)  is displayed in table \ref{multiscalar}.
 
\begin{table}[http!]
\renewcommand{\arraystretch}{1}
\begin{center}
\scalebox{1}[1]{
\begin{tabular}{|l|l|l|}
\hline
$N$ & {Type} & {Components} \\
\hline
 3 & {Hypergeometric} &     $\Lambda$ $\kappa$, $\Lambda$ m,~ $\kappa$ m,~$\kappa$ $\gamma$  \\
4 & {Heun} &   $\Lambda$ $\gamma$,~ m$\gamma$,~ $\Lambda$ $\kappa$ $\gamma$   \\
5 & {Gen.~ Heun} &    $\kappa$ m $\Lambda$,~   $\kappa$ m $\gamma$ \\
7 & {Gen.~ Heun} &    $\Lambda$   m $\gamma$,~  $\Lambda$ $\kappa$ m $\gamma$ \\
\hline
\end{tabular}
}
\end{center}
\caption{\it For each choice of fluid components, we display the number of singularities $N$ and the type of ODE governing scalar cosmological perturbations} 
\label{multiscalar}
\end{table}

The explicit forms of the `potentials' $Q(a)$ are given in appendix \ref{app:appendix}. 
  
\subsection{Tensor perturbations}
Let us now consider tensor perturbations of the simple (transverse traceless) form\footnote{This is what in GW community is called an $\times$ polarization, while a $+$ polarization with the same wave-vector in the third direction ${\bf k} = (0,0,k)$ would involve $dx_1^2 - dx_2^2$.}
 \be
 ds_4^2=-{da^2 \over a^2 b(a)^2} +  a^2 \left( dx_i^2+ h(a, x_3) dx_1 dx_2 \right) \ .
 \ee
 The linearized Einstein equations yield $\delta \rho=\delta \phi=0$. After making the \emph{Ansatz}
 \be
 h(a,x_3) \ = \ e^{-i k x_3}\,\hat{h}(a) \ ,
 \ee
 one finds the following ODE  
 \be
 \label{htteq}
2 a^4 \rho _0(a) \hat{h}''(a)+ a^3 \left(a \rho _0'(a)+8 \rho _0(a)\right) \hat{h}'(a)+ 6 M_{\text{Pl}}^2\,  k^2\, \hat{h} (a) =0 \ ,
 \ee
 The equation can be written in  the canonical form (\ref{eqcan}), with
 \be
 \hat{h}(a)={ \Psi(a) \over a^2 \rho(a)^{1\over 4}}
 \ee
 and
 \be
Q(a)= -\frac{4 a^4 \rho _0 \rho _0''-3 a^4 (\rho _0')^2+16 a^3 \rho _0 \rho _0'+32 a^2 \rho _0^2-48\,M_{\text{Pl}}^2\,   k^2 \rho _0}{16 a^4 \rho _0^2}
 \ee

 \subsubsection*{A single component universe}
 
  In the case of a single component universe, 
 \be
 \rho_0(a)=3 (M_{\text{Pl}}\, H_0)^2  a^{-3(1+w) }  
 \ee
  the linearized field equation for tensor  adiabatic perturbations (\ref{htteq})  becomes
  \be
  2\, a^{1-3 w}    \hat{h}''(a)- (3 w-5) a^{-3 w}   \hat{h}'(a)+2\, \ft{k^2}{H_0^2} \hat{h}(a)=0
  \ee
 that can be solved again in terms of Bessel functions
   \be
\label{sol_1Comp_w}
\hspace{-3mm} \hat{h}(a)\,=\, a^{\frac{3w-3}{4}} \left[c_1 \, J_{\alpha}\left(\gamma a^{1+3w\over 2} \right)+c_2 \, J_{-\alpha}\left( \gamma a^{1+3w\over 2} \right)\right]
 \ee
 with $c_1$, $c_2$ arbitrary real constants and  
 \be
 \alpha =\frac{3-3w}{2 (1+3w)} \  \qquad  , \qquad \gamma={ 2   k   \over   H_0  (1+3 w)} 
 \ee
 %Note that  $\alpha>0$  for $1>w>-1/3$ in this case, while $\alpha<0$  for $w>1$ and for $-1/3>w(>-1)$. As a consequence regularity conditions in the far past are subtler to impose for `exotic' fluids.
 
  \subsubsection*{A multi-component universe}
 
 Linear tensor perturbations around a multi-component  universe are described by the Schr\"odinger-like equation (\ref{eqcan}) with singularities at $a=0,\infty$ and at the zeros of $\rho_0(a)$. 
Very much as for scalar perturbations, the number of singularities depends on the specific choice of $\rho_0(a)$. For $\L$CDM  multi-component cosmologies with
\be 
\rho_0(a)  =  3 (M_{\text{Pl}}\, H_0)^2 \, \left(  \Omega_\Lambda + \Omega_{\text{m}} a^{-3} +\Omega_\gamma a^{-4}  \right)  
\ee
 the pattern of singularities  is displayed in table  \ref{multitens}.

\begin{table}[http!]
\renewcommand{\arraystretch}{1}
\begin{center}
\scalebox{1}[1]{
\begin{tabular}{|l|l|l|}
\hline
$N$ & {Type} & {Components} \\
\hline
4 & {Heun} &   $\Lambda$ $\gamma$,~ m$\gamma$ \\
5 & {Gen.~ Heun} &    $\Lambda$ m \\
7 & {Gen.~ Heun} &    $\Lambda$   m $\gamma$ \\
\hline
\end{tabular}
}
\end{center}
\caption{\it For each choice of fluid components, we display the number of singularities $N$ and the type of ODE governing tensor cosmological perturbations} 
\label{multitens}
\end{table}

 Interestingly, a comparison of Table \ref{multitens} against Table \ref{multiscalar} shows that tensor perturbations present a different analytic structure with respect to scalar perturbations. 
     The explicit forms of the `potentials' $Q_T(a)$ are given in appendix \ref{app:appendix} for various choices of multi-component fluids.

 \section{Cosmological Perturbations within the $\L$CDM model}
 
In this section we study cosmological perturbations around a flat 3-component universe by splitting the history in two 2-component phases separated by two transients. The first phase describes a universe evolving from an era dominated by radiation
to one dominated by matter. The second phase describes the evolution from matter to a de Sitter vacuum. We consider in turn scalar and tensor perturbations.

\subsection{Scalar perturbations}

The energy densities are given by
\bea
\gamma m: \qquad \rho_{\rm 0I} &=&     3 (M_{\text{Pl}}\, H_0)^2 \, \Omega_{\text{m}}\left(    a^{-3} +{\zeta}_{\rm I}  a^{-4}  \right)  \nn\\
m \Lambda: \qquad \rho_{\rm 0II} &=&  3 (M_{\text{Pl}}\, H_0)^2 \,\Omega_{\text{m}} \left(  {\zeta}_{\rm II} +   a^{-3}    \right)  
\eea 
with 
\be
{\zeta}_{\rm I}={\Omega_\gamma \over \Omega_m} \qquad ,\qquad {\zeta}_{\rm II}={\Omega_\Lambda\over \Omega_m} 
\ee
Adiabatic perturbations  around the corresponding backgrounds are described by  the second order ODE
  \bea
&&\Phi_{\rm I}  ''(a)
   \left(a  +{\zeta}_{\rm I} \right)+\frac{\Phi_{\rm I} '(a)
   \left(21 a^2  +54 a {\zeta}_{\rm I}   +32 {\zeta}_{\rm I} ^2\right)}{ 2a (3 a +4  {\zeta}_{\rm I} )}+ \frac{\Phi _{\rm I} (a) {\zeta}_{\rm I}  \left( \ft{4 a \hat{k}^2}{ \Omega_m H_0^2 } +3   \right)}{3a (3a +4  {\zeta}_{\rm I} )}=0 \nn\\
&&   a \Phi_{\rm II} ''(a) \left(a^3 \zeta_{\rm II}+ 1\right)+\Phi_{\rm II} '(a) \left(5 a^3 {\zeta}_{\rm II}+\frac{7  }{2}\right)+3 a^2 \Phi_{\rm II}
   (a) \zeta_{\rm II}=0 \label{eqradmat}
 \eea
 with
 \be
 \hat{k}={k\over \sqrt{\Omega_m} H_0}
 \ee
  The  first equation, for $k\neq 0$, is of  Heun type with singularities at $0,\infty,  -{\zeta}_{\rm I} ,-\ft43 {\zeta}_{\rm I} $.
 The second one is of  hypergeometric type. 
   The general solution of the latter can be written as
 \be
\Phi_{\rm II}(a)=  \sqrt{a^3 \zeta_{\rm II }+1}  \left[ \frac{2c_3}{5 }  \, _2F_1\left(\frac{5}{6},\frac{3}{2};\frac{11}{6};-a^3\zeta_{\rm II } \right)+ c_4 a^{-{5\over 2}} \right]
 \ee 
 We  build the solution by gluing local solutions in the two transients radiation-matter (I) and matter-vacuum (II) along the matter plateau, i.e. by defining
 \bea
 \Phi(a)=\left\{ 
\begin{array}{ccc}
 \Phi_{\rm I}(a) & ~~~~~  & a < a_0  \\
  \Phi_{\rm II}(a) & ~~~~~  & a>a_0  \\ 
\end{array}
\right.
 \eea
  with $a_0$ a point along the matter plateau.  We impose regular boundary conditions in the far past $a=0$, and the matching of the two solutions along the matter plateau
\bea
 \Phi_{\rm I} (a)  & \underset{a \to 0}{\approx}  & 1 \nn\\
 \lim_{a \to \infty }  \Phi_{\rm I} (a)  &=& \lim_{a \to 0}   \Phi_{\rm II} (a) \label{regmatch}
\eea

\subsection*{ k = 0 }

In the case of $k=0$, also the equation for the perturbation  in the $\gamma m$ transient can be solved in analytic terms leading to \cite{Kodama:1986fg}
\be
\Phi_I (a) = \frac{c_1 \sqrt{a+{\zeta}_{\rm I} }}{a^3}+\frac{2 c_2 \left(9 a^3+2 a^2 {\zeta}_{\rm I}-8 a {\zeta}_{\rm I}^2-16 {\zeta}_{\rm I}^3\right)}{15 a^3}
\ee
The  regularity and matching conditions (\ref{regmatch}) determine the coefficients $c_i$  to be
\be
c_1={8 \over 5}{\zeta}_{\rm I}^{5/2} \qquad ,\qquad c_2={3\over 4}  \qquad ,\qquad c_3={9\over 4} \qquad , \qquad c_4=0
\ee
At late times one finds
   \be
   \Phi_{\rm II} (a)   \underset{a\to \infty}{\approx}  {9 \Gamma(\ft23)\Gamma(\ft{11}{6}) \over 5 \sqrt{\pi} {\zeta}_{\rm II}^{1\over 3} a} +\ldots \approx {0.998\over a}+\ldots
  \ee
  The resulting function is almost identical to that plugged in Fig.~\ref{fig_sc}L for $\hat{k}$ slightly different from zero.  
  
  \subsection*{ k $\neq$ 0 }
  
  For $k\neq 0$, the equation in the domain I is of Heun type and cannot be solved in terms of elementary functions. The solution can be obtained by numerically integrating the differential equation imposing regular boundary conditions near $a=0$.  A qualitative description of the solutions can be obtained from a WKB analysis of the equation written in the Schrodinger like form (\ref{eqcan}). 
  We can distinguish two main regions according to the sign of $Q(a)$. For $k$ large enough,  $Q(a)$ is typically positive leading to an oscillatory behaviour of perturbations. 
  For $k$ zero or small, the perturbation $\Phi(a)$ follows the transient structure of the background, slightly delayed in time. 
   
    To estimate the critical value $\hat{k}_c$  where transitions between the two behaviours take place, we look for a zero of $Q(a)$ that it is also a zero of its derivative\footnote{Strictly speaking this exists only for small $a$.}.
  To be concrete, let us focus on early times $a\to 0$. Expanding $Q(a)$ for $a$ small one finds
  \be
  Q(a) \underset{a\to 0} {\approx} -{2 \over a^2} +{3\over 4 a \zeta_I } + { 64 \hat{k}^2 \zeta_I -117\over \zeta_I^2} +\ldots
  \ee
  The critical equations 
 \be
 Q(a_c)=Q'(a_c)=0
 \ee
are   solved by
\be
a_c={16 \zeta_I\over 3} \qquad, \qquad \hat{k}_c={3\over 8} \sqrt{23 \over 2 \zeta_I} 
\ee
 In Fig.~\ref{fig_sc}R  we display the numerical solution for $\Phi(a)$ for two representative choices of $\hat{k}$ below and beyond $\hat{k}_c$.  
 \begin{figure}[t]
  \begin{minipage}{\textwidth}
    \centering
 \includegraphics[width=0.48\textwidth]{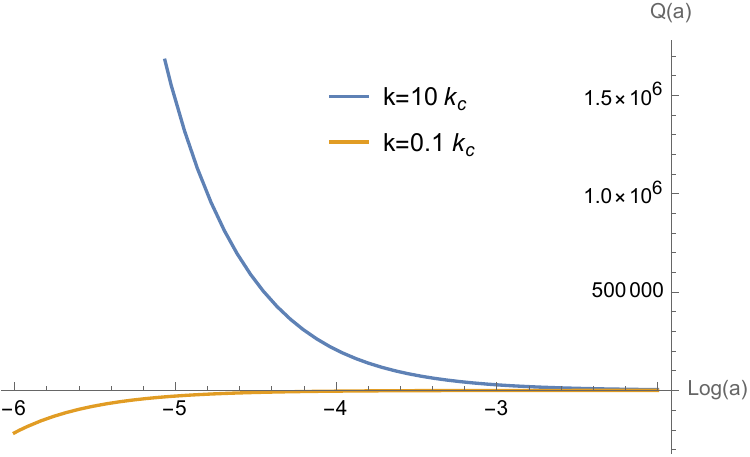}  \includegraphics[width=0.48\textwidth]{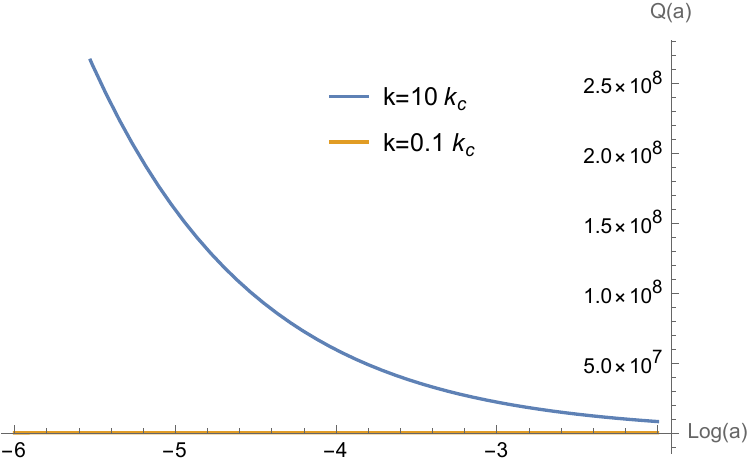} 
    \caption{\it   The `potential' $-Q(a)$ for two choices of $k$ above (blue) and below (orange) $k_c$:   L)  Scalar perturbations R)  Tensor perturbations }
    \label{fig_sc}
  \end{minipage}
\end{figure}

\begin{figure}[t]
  \begin{minipage}{\textwidth}
    \centering
 \includegraphics[width=0.45\textwidth]{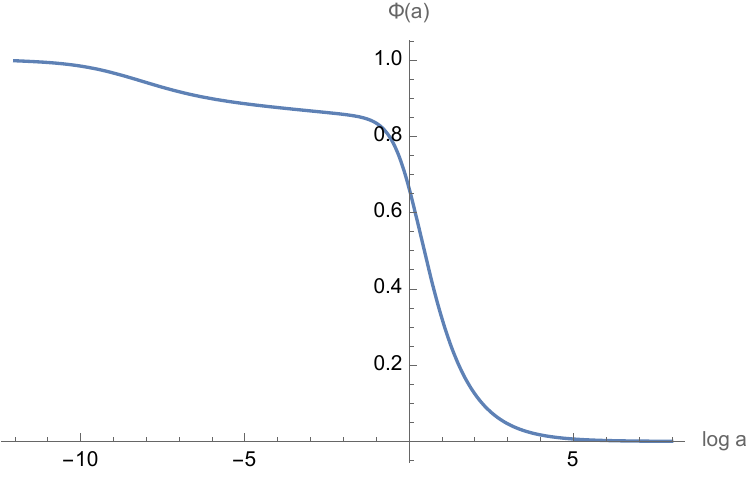} \includegraphics[width=0.45\textwidth]{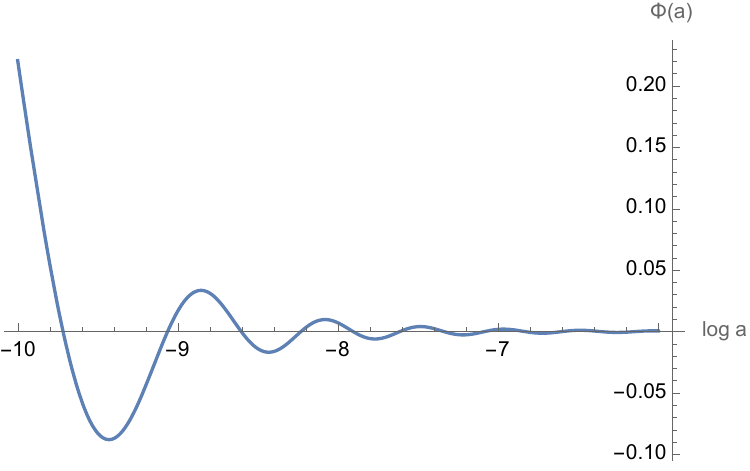}
    \caption{\it   Numerical solution for scalar perturbations $\Phi(a)$  L) $\hat{k}=0.1 \hat{k}_c$;  R) $\hat{k}=10 \hat{k}_c $. }
    \label{fig_sc}
  \end{minipage}
\end{figure}
  
 \subsection{Tensor perturbations}

\begin{figure}[t]
  \begin{minipage}{\textwidth}
    \centering
\includegraphics[width=0.45\textwidth]{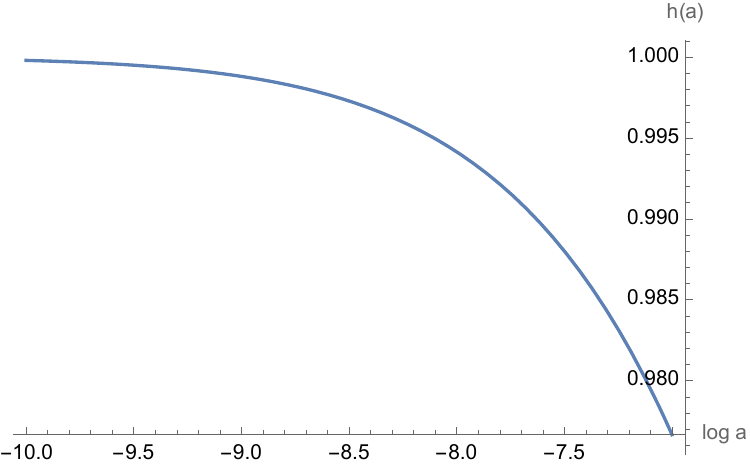} \includegraphics[width=0.45\textwidth]{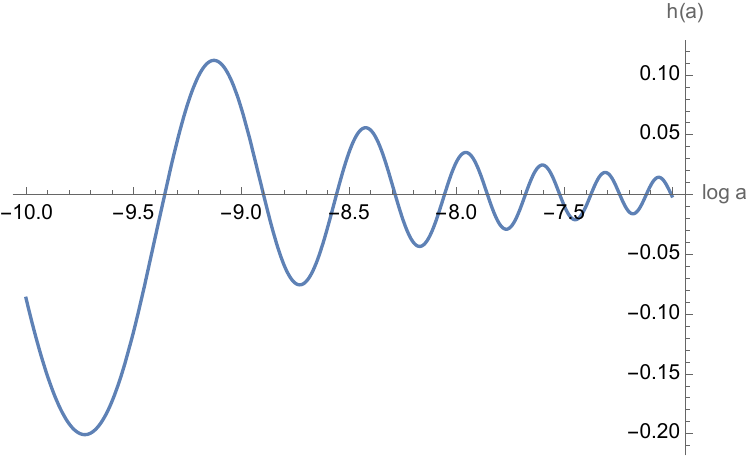}
    \caption{\it Numerical solution for  tensor perturbations $\hat{h}(a)$: L) $\hat{k}=0.1 \hat{k}_c$;  R) $\hat{k}=10 \hat{k}_c$. }
    \label{fig_ten}
  \end{minipage}
\end{figure}

 The study of tensor perturbations for our $\Lambda$CDM model proceed {\it mutatis mutandis} along the same steps as before. The equations describing the radiation-matter and matter-vacuum phases 
 are now    
 \bea
2a\left(a+{\zeta}_{\rm I}\right)  \hat{h}_{\rm I} ''(a)+\left(5 a+4{\zeta}_{\rm I}\right)   \hat{h}_{\rm I}  '(a) + \ft{2a k^2}{\Omega_m H_0^2} \hat{h}_{\rm I}  (a)=0\nn\\
2a \left(a^3 {\zeta}_{\rm II}+1\right)   \hat{h}_{\rm II} ''(a)+ \left(8 a^3 {\zeta}_{\rm II}+5\right)   \hat{h}_{\rm II} '(a)+\ft{2k^2}{\Omega_m H_0^2} \hat{h}_{\rm II} (a)=0
 \eea
 The first equation is a confluent Heun equation while the second one has five singular points.

 \subsection*{k = 0}
 
 Again for $k=0$ the equations can be solved in terms of elementary functions
 \bea
 \hat{h}_{\rm I} &=& c_1 +c_2\left[  \arctanh\sqrt{1+a {\zeta}_{\rm I}^{-1}}  - { {\zeta}_{\rm I} \over a} \sqrt{ 1 +a{\zeta}_{\rm I}^{-1} } \right] \nn\\
 \hat{h}_{\rm II} &=& c_3 +c_4 a^{-{3\over 2}} \sqrt{1+a^3 {\zeta}_{\rm II}}
 \eea
but now the only solution satisfying the gluing conditions  (\ref{regmatch})  is the constant $\hat{h}(a)=1$. 
 
 \subsection*{k $\neq $ 0 }

For $k \neq 0$, the profile $\hat{h}(a)$ can be found by numerically  integrating 
 the differential equation. Again the solutions  exhibit an oscillatory behaviour  for $\hat{k}$ beyond a critical value and a monotonous fall for $\hat{k}$ small. 
 Now the Schrodinger like potential at early times is given by
 \be
Q(a) \underset{a\to 0}{\approx} -\frac{1}{2 a \zeta _{\rm I}}+\frac{a \left(-8 \zeta _{\rm I} \hat{k}^2-7\right)}{8 \zeta _{\rm I}^3}+\frac{16 \zeta _{\rm I} k^2+11}{16 \zeta _{\rm I} }+\ldots
 \ee
 and the critical point is
 \be
 a_c= \ft{2}{3} \left(\sqrt{22}-4\right) \zeta _{\rm I}  \qquad ,\qquad  \hat{k}_c = \frac{\sqrt{5+4 \sqrt{22}}}{4 \sqrt{\zeta _1}}
 \ee
 The solutions for some representative choices of $\hat{k}$ below and beyond $\hat{k}_c$ are plotted in Fig.~\ref{fig_ten}.

\section{Seiberg-Witten / cosmology correspondence}

 \begin{figure}[t]
    \centering
    \includegraphics[width=0.7\textwidth]{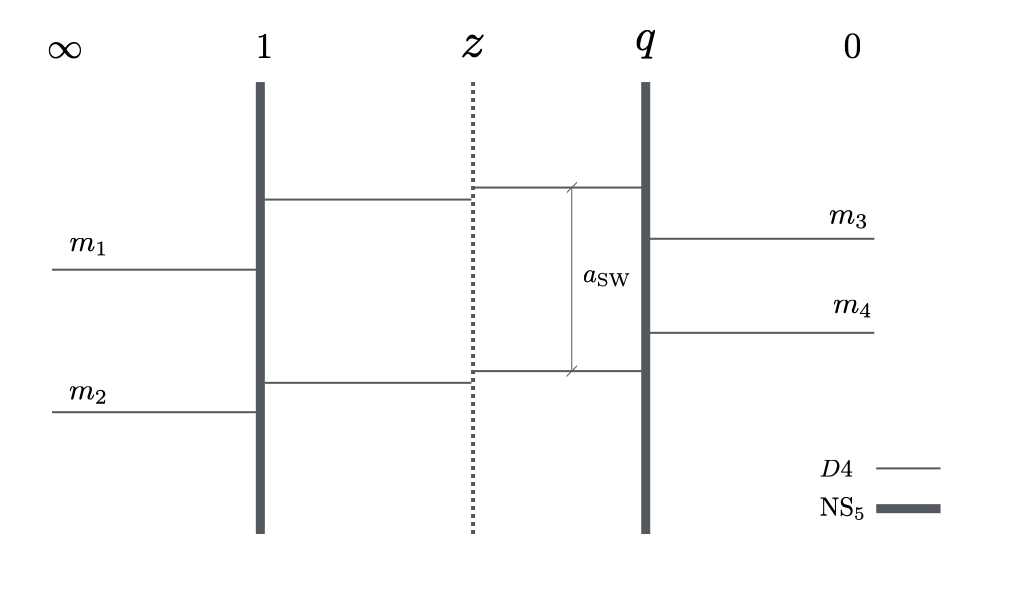}
    \caption{\it Brane realization of the $SU(2)^2$ quiver gauge theory. }
    \label{figHW}
\end{figure}
 
In this section we briefly review the quantum SW / gravity correspondence relating partition functions of linear quiver gauge theories with ${\cal N}=2$ supersymmetry  to solutions of generalized Heun equations that describe cosmological perturbations, discussed so far. 
 
 In string theory,  linear quiver gauge theories with ${\cal N}=2$ supersymmetry in four dimensions can be realised, following  Hanany and Witten  \cite{Hanany:1996ie}, by suspending parallel  D4-branes between stacks of NS5-branes  distributed along a line and considering the fluctuations of open strings connecting D4-branes between themselves.  
 Let us denote by $z_i$ with $i=1,2,\ldots N$, the positions of NS5 brane.  Open strings connecting D4 branes inside a finite interval along the $z$ direction realise gauge degrees of freedom (vector multiplets), while those connecting D4-branes belonging to two adjacent intervals gives rise to bi-fundamental matter (hyper-multiplets). The gauge couplings are specified by the ratio $q_i=z_{i+1}/z_i$ of the starting and ending point of the gauge D4-brane while the positions of D4-branes along the vertical direction parametrise scalar vev and masses. The system can be lifted to M-theory and realised in terms of a single M5-brane wrapping a two-dimensional Riemann surface, {\it a SW complex curve}. The rank of the gauge groups in the individual nodes are associated to the order of the differential equation, so for our purposes, $SU(2)$ gauge group are enough to capture the relevant dynamics.
 
 Let us start and consider the case of an equation of Heun type.  In Fig.~\ref{figHW} we display  the corresponding brane system that realizes an $SU(2)^2$ quiver gauge theory that consists of pairs of D4-branes suspended in between five NS5-brane located at positions $z_i= (\infty,1,z,q,0)$. D4-branes starting at $z_1=\infty$ or $z_4=0$ have zero coupling $q_1=q_4=0$, so they can be viewed as flavour branes. The two $SU(2)$ gauge groups arise from D4-branes extending along the $[1,z]$ and $[z,q]$ intervals.  The system we are interested in here involves  the presence of a very light NS5-stack, that we indicate with a dotted vertical line. D4-branes at the two sides of this stack 
 tend to align with one another otherwise the tension at the two sides will break the light separation wall.  Indeed, in first approximation, the presence of this extra stack can be ignored leaving only three D4-brane stacks  realising an $SU(2)$ gauge theory with four flavours.  
The geometry is captured by the ``classical'' SW curve  
   \be
  (x,z)\in \mathbb{C}^2 :\qquad   P_0(x) z^2-P_1(x) z+q P_2(x)  =0 \label{swcurve} 
   \ee  
   with $q=e^{2\pi i \tau}$ the (complexified) gauge coupling and $P_{n}(x)$ some polynomials of degree two specifying the positions of the D4-branes inside each interval (masses and vev's) . To make contact with gravity, we consider the gauge theory on a curved
 Nekrasov-Shatashvili (NS) $\Omega$-background \cite{Nekrasov:2002qd, Flume:2002az, Bruzzo:2002xf, Nekrasov:2009rc, Poghossian:2010pn, Fucito:2011pn}.  The net effect of this background is to make the curve ``quantum", promoting $(x,z)$  to non-commuting operators, such that $x=-z\partial_z$. 
   Since $P_{n}(x)$ are polynomials of degree two,  the curve (\ref{swcurve}) translate into  an ordinary differential equation of second order with four singularities $z_i=\infty,1,q,0$, {\it i.e.} an equation of Heun type. Moreover, the solution of the differential equation can be related to the partition function of the $SU(2)^2$ quiver gauge theory obtained by adding the extra  NS5-brane at position $z$.    
  
 A similar correspondence can be built  for equations of generalized Heun type, that we have shown to govern cosmological perturbations around general FLRW backgrounds. One simply adds a heavy NS5-stack, {\it i.e.} a gauge group factor, for each extra singularity and establishes a dictionary between gauge theory variables and cosmological ones. In particular the chosen time variable $a$ is related to the variable $z$ parametrizing the position of the `light' NS5-brane, see Fig.\ref{figHW}. Notice that a crucial role is played by the quantum SW period ${\mathfrak a}$, not to be confused with the scale factor $a$. In the similar context of BH or fuzzball perturbation theory ${\mathfrak a}$ plays the role of `renormalized' angular momentum, denoted by $\nu$ in  \cite{Mino:1997bx, Sasaki:2003xr}.  In \cite{Fucito:2023afe, Bianchi:2024vmi} it is shown that ${\mathfrak a}=\nu +{1\over 2} \sim \ell +{1\over 2}$.  
 
 The generalized Heun function $\Psi$ can be written as a multi-instanton series, {\it the  Nekrasov partition function} of the quiver, in the limit where all couplings $q_i$ are small.  Connection formulae relating solutions around two singular points in the $z$-plane can be derived using braiding and fusion relations, descending from crossing symmetry of the dual two-dimensional CFT according to Alday-Gaiotto-Tachikawa (AGT) duality \cite{Belavin:1984vu,Alday:2009aq,Nekrasov:2009rc}. 
    
     \subsection{The SW gravity  Heun correspondence}

For concreteness let us focus on the case of cosmological perturbations described by an equation of Heun type. The corresponding gauge theory description is given in terms of a SU(2) gauge theory with $N_f=(N_0,N_2)$ flavours,  $N_0=N_2=2$ counting the number of flavour D4 branes (Left and Right) in the brane setup. 
 The quantum SW curve describing the low energy dynamics of ${\cal N}=2$ supersymmetric $SU(2)$ gauge theory with four flavours on a NS $ \Omega$-background  is given by
the quantum version of (\ref{swcurve}) where $x$ and $z$ are promote to non-commuting operators.  The quantum curve can be written, either as a difference equation 
\be
P_0(x+\ft12 ) y(x) y(x-1)-P_1(x) y(x-1) +q  P_2(x-\ft12 ) =0  
\ee
by thinking of $z=e^{-\partial_x}$, or as a differential equation
 \be
 \left[    P_0(- z\partial_z+\ft12  ) - P_1(-z\partial_z  ) z^{-1}+q P_2( -z\partial_z-\ft12 ) z^{-2}\right] W(z)=0 \label{qswcurve}
 \ee 
   with $x=-z\partial_z$ and
  \bea
 P_0(x) &=&   (x-m_1)(x-m_2) \qquad , \qquad  P_2(z)= (x-m_3)(x-m_4) \nn\\
 P_1(x)&=& x^2-u+q\left(  x^2+u +\ft12 -(x+\ft12)\sum_i m_i +\sum_{i<j} m_i m_j \right)
 \eea
    Here $u = \langle {\rm Tr}\phi^2\rangle$ parametrises  the Coulomb branch of the moduli space and $m_f$ the masses of the hypers.
%We introduce the superfluous parameter $\delta=0,1$ for later convenience, but we stress that only the product $P_L(x)P_R(x)$ is physically relevant. 
     Equation (\ref{qswcurve}) can always be written in the form (\ref{eqcan}) by taking
  \be   
     W(z) = z^{1-{m_3+m_4\over 2}} (1-z)^{-{m_1+m_2+1\over 2}}  (z-q)^{{m_3+m_4-1\over 2}} \Psi(z)  
   \ee  
   leading to
   \bea
   Q_{22} (z) &=& \frac{1- 
   \left({m_1+m_2}\right)^2}{4(z-1)^2}+\frac{1- 
   \left({m_3-m_4}\right)^2}{4z^2}+\frac{1- \left({m_3+m_4}\right)^2}{4(z-q)^2}\nn\\
  &&\qquad   +\frac{2 m_1 m_2+m_3^2+m_4^2-1}{2 (z-1)
   z}+\frac{(1-q)U
 }{(z-1) z (z-q)} \label{Q22}
   \eea
where the subscript $22$ relates to the HW brane setup with $N_f=(2,2)$ and
   \be
   U= u+\ft14 -\ft{1}{2} \left(m_3^2+m_4^2\right)
   -\frac{q \left(1-m_1-m_2\right) \left(1-m_3-m_4\right)}{2
   (1-q)}
   \ee
    We will also need the reduced confluent limit, whereby two hyper-multiplets decouple from the dynamics, defined by taking $q\to 0$, $m_3,m_4\to \infty$, keeping fix the product $q m_3 m_4 = q_2$.    In this limit one finds
   \be
  Q_{20}(z)={-}\frac{q_2}{z^3} {+}\frac{1{-}4 u}{4 z^2} {+} \frac{1{-}4 m_1 m_2{-}4 u}{4 z}{+}\frac{4 m_1 m_2{+}4 u{-}1}{4 (z-1)}{+}\frac{1 {-}(m_1{+}m_2)^2 }{4 (z-1)^2}\label{q20}
   \ee
   where the subscript $20$ relates to the HW brane setup with $N_f=(2,0)$. 
   
    The difference equation can be solved for $y(x)$ or $y(x-1)$ by infinite continuous fractions given in terms of $P_i(x)$. Equating the two expressions one finds the consistency condition  \cite{Poghosyan:2020zzg}
\bea
\label{fractionequality}
\frac{ q M( \mathfrak{a}+1)}{P_1(\mathfrak{a}+1)-\frac{ q M(\mathfrak{a}+2)}{P_1(\mathfrak{a}+2)-\ldots%\frac{ q M(a+3\hbar)}{P_2(a+3\hbar)-\,\cdots}
}}
+\frac{ q M(\mathfrak{a})}{P_1(\mathfrak{a}-1)-\frac{ q M(\mathfrak{a}-1)}{P_1(\mathfrak{a}-2)-\ldots %\frac{ q M(a-2\hbar)}{P_2(a-3\hbar)-\,\cdots}
}}=P_1(\mathfrak{a})
\eea
where, for the sake of simplicity, we used the notation
\be
M(x)=P_0(x-1)P_2(x)=\prod_{i=1}^4 (x-m_i -\ft12)
\ee
  The solution $\mathfrak{a}(u)$ of (\ref{fractionequality}) gives the quantum SW period and will play a crucial role in what follows. It can be computed perturbatively, order by order in $q$,
  leading to 
  \be
 \mathfrak{a}(u)= \sqrt{u} + {q\over 2\sqrt{u} } \Bigg[ {4 u {\,+\,}3  \over 8} {+}  {2 m_1 m_2 m_3 m_4 \over 4 u{-}1}  {-} \ft{1}{2} \sum_i m_i  {+} \ft12 \sum_{i<j} m_i m_j \Bigg] +O(q^2) \nn\\
  \ee
Notice that in the large $u$ semi-classical limit $\mathfrak{a}(u) \sim \sqrt{u}$.
 
  \subsection{The solutions $\Psi_\alpha(z)$ }
  
   The solutions $\Psi_\alpha(z)$ to the generalized Heun equation can be related to the partition function of an $SU(2)^2$ quiver gauge theory obtained by introducing the light NS5 stack in an $SU(2)$ system. 
  The position $z$ of the light NS5 stack determines the region of spacetime described by the quiver partition function. For example, the partition function describing the solution in the region $1 \gg z \gg q $ is computed by means of the localization  formula  
  \bea
\Psi_\alpha(z)  =  \lim_{b \to 0}\,   \,  z^{{1\over 2} +\alpha \mathfrak{a}}     \,
\left(1{-} z  \right)^{ 2k_{0}{-}1   \over 2 b^2  } \left( 1- {q\over z} \right)^{{1\over 2}+k_2}   
	{ Z_{\rm inst}{}_{p_0}{}^{k_0}  {}_{\mathfrak{a}^{-\alpha}}{}^{ k_{\rm deg}}   {}_{\mathfrak{a}}{}^{k_2}{}_{p_3}  (z,\ft{q}{z} ) \over
	Z_{\rm inst}{}_{p_0}{}^{k_0} {}_{\mathfrak{a}}  {}^{k_2}{}_{p_3} (q)   }  \label{psidef}
\eea
with
{\small
\bea
Z_{\rm inst}{}_{p_0}{}^{k_0} {}_{\mathfrak{a}}{}^{k_2}{}_{p_3}  (q) &=&  
	 \sum_{ W}  q^{ |W |}
	 \frac{ z^{\rm bifund}_{\emptyset,W} (p_0,\mathfrak{a},-k_0)   z^{\rm bifund}_{\emptyset,W} (\mathfrak{a},p_3,-k_2) 
	 }{     z^{\rm bifund}_{W,W} (\mathfrak{a},\mathfrak{a},\ft{b^2+1}{2}) }   \label{zinstG2} \\
	Z_{\rm inst}{}_{p_0}{}^{k_0}  {}_ {p_1 } {}^{k_1}  {}_{p_2}{}^{k_2}{}_{p_3} (q_1,q_2) &=&
	 \sum_{ Y_1, Y_2}  q_1^{  |Y_1| } q_2^{ |Y_2 |}
	 \frac{  \prod_{i=0}^2 z^{\rm bifund}_{Y_i,Y_{i+1} } (p_i, p_{i+1},-k_i) }{ \prod_{i=1}^2 z^{\rm bifund}_{Y_i,Y_i} ( p_i , p_i,\ft{b^2+1}{2})    }
	\label{inst1}\eea
	}
 the instanton partition functions of the involved $SU(2)$ and $SU(2)^2$ gauge theories, with and without the extra light NS5 stack. 
 The sums in (\ref{inst1}) run over pairs of Young tableau $\{Y_{1\pm} \}$, $\{Y_{2\pm} \}$ while $|Y_i|$ denote the total number of boxes in each pair and $Y_0=Y_3=\emptyset$.
The functions  $z^{\rm bifund}_{Y,W }$ represent the contributions of a hyper-multiplet transforming in the bi-fundamental representation of the gauge group and are given by a product over the boxes of the Young tableau  
\bea
	 z^{\rm bifund}_{\Lambda, \Lambda' }( p , p' , m ) &=&   \prod_{\beta,\beta' =\pm}
	    \prod_{(i,j)\in \Lambda_\beta}  \left[  E_{\Lambda_\beta,\Lambda'_{\beta'}}(\beta p{-}\beta' p',i,j)  {-} m  \right]  \nn\\
	    && \times  \prod_{(i',j')\in \Lambda'_{\beta'}} \left[  { -}E_{\Lambda'_{\beta'},\Lambda_\beta}(\beta' p' {-}\beta p,i',j')  {-} m  \right]  
\label{inst2}
\eea	
with
\bea
E_{\Lambda, \Lambda'}(x,i,j) &=& x- (\lambda_{ \Lambda' j}^T-i) + b^2 (\lambda_{ \Lambda i}-j+1)-\ft{b^2+1}{2}
\label{inst3}
\eea
where $\lambda_{ \Lambda i}$ is the number of boxes in the $i$-th row of the tableau $\Lambda$ and $\lambda_{ \Lambda' j}^T$ is the number of boxes in the $j$-th column of the tableau $\Lambda'$. Finally
\bea
p_0&=&{m_1-m_2\over 2}   \quad, \quad k_0={m_1+m_2\over 2} \quad, \quad p_3={m_3-m_4\over 2} \quad, \quad k_2={m_3-m_4\over 2} \nn\\
k_1&=&  k_{\rm deg}=\ft12+b^2 \quad , \quad p_1=\mathfrak{a}^\alpha=\mathfrak{a}+\alpha \ft{b^2}{2}  \quad, \quad p_2=\mathfrak{a}
\eea
  We have to impose regular boundary conditions at early times on the wave function
  \be
  W(z)   \underset{z\to \infty}{\sim}  z^{-{m_1+m_2+1\over 2}}  \Psi(z)   
  \ee
 so we need a connection formula relating $ \left\{ \Psi_\alpha \right\} $  to the basis  $ \left\{ \widetilde{\Psi}_\alpha \right\} $ of solutions in the domain where $z \gg 1 \gg q$. 
   The connection matrix $B_{\alpha'' \alpha}$ can be derived from crossing symmetry in the AGT dual conformal field theory
 and can be pictorially represented as
 \bea
 \label{braid_relation}
 \widetilde{\Psi}_{\alpha''} (z) =
\begin{tikzpicture}[baseline={(current bounding box.center)}, node distance=0.8cm and 0.8cm]
\coordinate[label=above:$k_{\rm deg}$] (k0);
\coordinate[below=of k0] (s0);
\coordinate[left=of s0] (p0);
\coordinate[right=1cm of s0] (s1);
\coordinate[above=of s1,label=above:$k_0$] (k1);
\coordinate[right=of s1] (p2);

\draw[line,dashed] (k0) -- node[label={[xshift=0.2cm, yshift=0.1cm]left:\scriptsize{$ z$}}] {} (s0);
\draw[line] (s0) -- (p0);
\draw[line] (s1) --node[label={[xshift=0.2cm, yshift=0.1cm]left:\scriptsize{$ 1$}}] {}  (k1);
\draw[line] (s1) -- (p2);
\draw[line] (s0) -- node[label={[yshift=0.1cm]below:$p_0^{\alpha''}$}] {} (s1);
\draw[line] (s0) -- node[label=below:$p_0$] {} (p0);
\draw[line] (s1) -- node[label=below:$\mathfrak{a}$] {} (p2);

\coordinate[right=of p2] (s2);
\draw[line] (p2) -- node[label={[yshift=0.15cm]below:$p_3$}] {} (s2);
\coordinate[above=of p2,label=above:$k_2$] (k2);
\draw[line] (p2) --node[label={[xshift=0.2cm, yshift=0.1cm]left:\scriptsize{$ q$}}] {}  (k2);

\end{tikzpicture}  \cdot \cdot \cdot &=& \sum_{\alpha}B_{\alpha'' \alpha} 
\begin{tikzpicture}[baseline={(current bounding box.center)}, node distance=0.8cm and 0.8cm]
\coordinate[label=above:$k_0$] (k0);
\coordinate[below=of k0] (s0);
\coordinate[left=of s0] (p0);
\coordinate[right=1cm of s0] (s1);
\coordinate[above=of s1,label=above:$k_{\rm deg}$] (k1);
\coordinate[right=of s1] (p2);
\draw[line] (k0) -- (s0);
\draw[line,dashed] (s0) --node[label={[xshift=0.2cm, yshift=0.1cm]left:\scriptsize{$ 1$}}] {}  (k0);
\draw[line] (s0) -- (p0);
\draw[line,dashed] (k1) -- node[label={[xshift=0.2cm, yshift=0.1cm]left:\scriptsize{$ z$}}] {} (s1);
\draw[line] (s1) -- (p2);
\draw[line] (s0) -- node[label={[yshift=0.15cm]below:$\mathfrak{a}^{-\alpha}$}] {} (s1);
\draw[line] (s0) -- node[label=below:$p_0$] {} (p0);
\draw[line] (s1) -- node[label=below:$\mathfrak{a}$] {} (p2);

\coordinate[right=of p2] (s2);
\draw[line] (p2) -- node[label={[yshift=0.15cm]below:$p_3$}] {} (s2);
\coordinate[above=of p2,label=above:$k_2$] (k2);
\draw[line] (p2) --node[label={[xshift=0.2cm, yshift=0.1cm]left:\scriptsize{$ q$}}] {}  (k2);

\end{tikzpicture} \cdot \cdot \cdot =\sum_{\alpha}B_{\alpha'' \alpha}  \Psi_{\alpha} (z) \nn\\
\eea
with
\be
B_{\alpha'' \alpha}   = \frac{e^{i \pi   \left( \alpha'' p_0- \alpha\mathfrak{a} {+}{b^2\over 2}\right)}\Gamma \left(1-2  \alpha'' p_0\right) \Gamma \left(-2  \alpha  \mathfrak{a} \right)}
{\Gamma \left(\frac{1}{2} -\alpha'' p_0-\alpha \mathfrak{a}+k_0 \right)
	\Gamma \left(\frac{1}{2}-\alpha'' p_0-\alpha \mathfrak{a}-k_0 \right)}  \label{braidingB}  
	\ee
The regular solution at $z=\infty$ is $\Psi(z)= \widetilde{\Psi}_-(z) $, so one finds
\be
\Psi(z) = \sum_{\alpha =\pm}B_{- \alpha} \Psi_{\alpha} (z) \label{psifinal}
\ee
  It is important to observe that the connection matrix $B_{\alpha'' \alpha}$  is nothing but the standard connection matrix of hypergeometric functions after the replacement of $\sqrt{u}$ with the full quantum SW period $\mathfrak{a}(u)$. To see this, we  first observe that setting $q\to 0$ in (\ref{Q22}) leads to a hypergeometric equation. Assuming $m_1>m_2$, the regular solution at large $z$ is
  \be
\Psi(z)=  \widetilde{\Psi}_- (z) = z^{m_1} (1-z)^{1-m_1 -m_2\over 2} \, {}_2F_1\left( \ft12{ -}m_1{-}\sqrt{u} , \ft12 {-}m_1{+}\sqrt{u};1{-}m_1{+}m_2;\ft{1}{z}\right) \label{tildepsiH}
  \ee
  with $ \widetilde{\Psi}_\pm$ related to each other by exchanging $m_1 \leftrightarrow m_2$. Using the hypergeometric connection formulas, one can rewrite 
   (\ref{tildepsiH}) in the form
  \bea
  \Psi (z) &=&    \sum_{\alpha =\pm}  e^{-i \pi  \left(\frac{1}{2} -m_1+\alpha \sqrt{u}\right)} \ft{\Gamma \left(1{-}m_1{+}m_2\right)  \Gamma \left({-}2  \alpha\sqrt{u} \right) 
  }{\Gamma \left(\ft12 {-}\alpha\sqrt{u} 
   {-}m_1 \right) \Gamma \left(\ft12 -\alpha \sqrt{u} {+}m_2 \right)}  \Psi_\alpha(z)  \label{psiq0}
  \eea
  with 
    \be
  \Psi_\alpha (z)\underset{q\to 0}{\approx} z^{{1\over 2} +\alpha \sqrt{u}} (1-z)^{1{-}m_1 {-} m_2\over 2} {}_2 F_1(\ft12{ -}m_1{+}\alpha \sqrt{u},\ft12{ -}m_2 {+}\alpha \sqrt{u},1{+}2\sqrt{u},z) 
  \ee
  It is easy to check that the coefficients in (\ref{psiq0})  are given by $B_{- \alpha}$ defined in (\ref{braidingB}) after replacing $\mathfrak{a} \to  \sqrt{u}$. 
  When $q \neq 0$, the hypergeometric functions are replaced by Heun functions given by the double instanton series (\ref{psidef}) and the connection matrix is given by
  (\ref{braidingB}) with  $\mathfrak{a}$ the quantum SW period.

   \subsection{Scalar and tensor perturbations}

     Now let us consider adiabatic perturbations of an expanding universe filled in with radiation and matter.  We introduce the short-hand notation
     \be
     \zeta = {\Omega_\gamma \over \Omega_m}  \qquad, \qquad \hat{k}={k\over H_0 \sqrt{\Omega_m} } 
     \ee
      Scalar  perturbations are described by a Heun equation that can be written in the Schr\"odinger like form (\ref{eqcan}) 
    with 
     \be
  Q(a)=   \frac{64 a^2 \zeta  k^2 \left(3 a^2{+}7 a \zeta {+}4 \zeta ^2\right){-}3 \left(189 a^4{+}924 a^3 \zeta {+}1820 a^2 \zeta ^2{+}1600 a \zeta ^3{+}512 \zeta ^4\right)   }{48 a^2 (a{+}\zeta )^2 (3 a{+}4 \zeta )^2   }
     \ee
 Comparing against (\ref{Q22}) one finds the gauge cosmology dictionary
 \bea
z &=& -\zeta\,  a^{-1}  \qquad , \qquad q = \frac{3}{4} \qquad , \qquad   u= \frac{4 \hat{k}^2 \zeta^2 }{3  }+\frac{33}{16}  \nn\\
 m_1&=& \frac{7}{4} \qquad , \qquad m_2= -\frac{5}{4} \qquad , \qquad  m_{3,4}=1\pm  \ft{1}{12} \sqrt{225 -64 \hat{k}^2 \zeta } 
\eea
Notice that $q = \frac{3}{4}$ is independent of $k$ since the $\beta$ function of the relevant gauge theory (with $N_f=4$) is zero. Indeed, as shown in \cite{Bianchi:2021yqs,Bianchi:2022wku,Bianchi:2022qph}, in general $q\sim (\omega M)^\beta$ where $M$ is some relevant mass scale and $\omega$ the frequency. Anyway, even for finite and non-tuneable $q$, we expect convergence of the instanton sum.
 
 On the other hand for tensor perturbations one finds the Schr\"odinger-like potential
 \be
 Q(a) =\frac{ \hat{k}^2}{(a+\zeta )}-\frac{5 a+8 \zeta }{16 a (a+\zeta )^2}
 \ee
 leading to a reduced confluent Heun equation that can be matched to $Q_{20}$ given in (\ref{q20}) after the identifications
 \be
 z=-\zeta \, a^{-1} \quad, \quad q_2=\hat{k}^2 \zeta \quad ,\quad m_1=\ft34 \quad , \quad m_2=-\ft14 \quad, \quad  u=\hat{k}^2 \zeta +\ft{9}{16} 
 \ee
 We notice that now  $q_2\sim k^2$  since $\beta =2$ in the $N_f=(2,0)$ case, and it can be tuned with  $\hat{k}^2$ to be arbitrarily small.
 
In both cases $\mathfrak{a} \approx \sqrt{u} + ... \sim k$ for large $k$. 
 
 \section{Conclusions}
 
 Let us briefly summarize the results of our present investigation and outline directions for future research. 
 
 First of all we have shown that choosing the scale factor $a$ as time coordinate\footnote{We thank Misao Sasaki very much for suggesting to change from our original choice of time variable $\tau = 1/a$ to $a$ itself. The variable $\tau$ looked more suitable for the late-time De Sitter phase but obscured some of the results that obtain neatly in terms of $a$.} allows to explicitly solve Einstein equations for a general FLRW model coupled to a multi-component perfect fluid. In particular we have written down the analytic form of the Hubble rate $b=H(a)$ and of the deceleration parameter $q(a)$ in Section 2.
 
 We have then analyzed linear perturbations around a general FLRW cosmological model. In Section 3 we have written down the relevant wave equations for both scalar and tensor adiabatic perturbations. After checking consistency with known results for a single component model, whereby the dynamics is governed by Bessel equations, we have shown that the general form of the linear perturbation equations is of the Heun type (with four regular singular points) or its generalizations, also in agreement with some results in the literature though with a different choice of time coordinate \cite{Kodama:1984ziu}.  We have also noticed that the singularity structure for scalar and tensor adiabatic perturbations may be rather different. 
 
We have largely neglected entropy or iso-curvature modes, that are known to couple to the adiabatic modes, we have focussed on, in the case of a multi-component fluid \cite{Kodama:1984ziu, Mukhanov:1990me, Vittorio:2017foh}. Since this coupling introduces non-homogeneous source terms in the equations for the adiabatic modes, the exact knowledge of the solutions of the homogeneous equations for purely adiabatic modes that we have achieved may not be sufficient to describe the actual behavior of an adiabatic perturbation. In order to estimate
this effect, one should know the strength of the coupling to entropy and iso-curvature modes, which would require determining the first non-linear corrections to the linear homogeneous equations and may be rather model dependent. Yet, treating the coupling as a non-linear correction allows to resort to the Green function method that would benefit a lot from our knowledge of the analytic solutions to the homogeneous equations. We hope to come back to this point.

 In Section 4 we have focussed on the $\Lambda$CDM model, consisting of three eras (radiation dominated, matter dominated and vacuum energy dominated) separated by two transients (radiation to matter, matter to vacuum energy). During the first transient, dynamics is governed by a Heun equation with four regular singularities: two `physical' for $a=0$ (far past) and $a=\infty$ (late time) and two `unphysical' at $a = - \zeta_I,-4\zeta_I/3$ with $\zeta_I =  \Omega_\gamma/\Omega_{\rm m}$. The second transient, from matter to vacuum energy, is instead governed by a simpler Hypergeometric equation. Imposing regularity and matching conditions we find the full solution by numeric integration of the differential equations. 
 
Finally we have related the ODE's for cosmological perturbations to the quantum Seiberg-Witten curves for ${\cal N}=2$ quiver gauge theories, proposing a new SW / cosmology correspondence. Very much as for BHs, fuzzballs and ECOs perturbations, one needs no more that linear $SU(2)$ gauge quivers with the number of nodes determined by the number of singularities. In particular we have spelled out the dictionary for a universe filled in with matter and radiation, whereby scalar perturbations are governed by a Heun equation corresponding to the qSW curve for SU(2) gauge theory with $N_f=(2,2)$ flavours (in the Hanany-Witten setup), while tensor perturbations are governed by a reduced confluent Heun equation corresponding to the qSW curve with $N_f=(2,0)$ matter. 
The solutions of the Heun equations are computed by the instanton partition functions of the quiver gauge theory. Thanks to the AGT correspondence the ODEs may be viewed as a conformal Ward identity for the conformal blocks of a Liouville-like CFT, and connection formulae, relating the behaviour of the wave function at early and late times, are derived from crossing symmetry.

Given our knowledge of the connection formulae for generalized Heun equations, we plan to further exploit the new gauge / cosmology correspondence in order to analyze the power spectrum and other observables along the lines of holographic cosmology approach, pioneered by Skenderis and Townsend \cite{Skenderis:2006jq,Skenderis:2007sm} and further elaborated on in \cite{McFadden:2009fg,Bzowski:2023nef}.

The very same gauge / cosmology correspondence can prove useful in analyzing inflation and pre-Big-Bang cosmology after the inclusion of additional (scalar) fields that drive inflation and contribute to the generation of primordial fluctuations and their non-gaussianity \cite{Sleight:2019hfp,Sleight:2020obc,Arkani-Hamed:2023kig}.  

Another interesting direction for future investigation is the mass spectrum of Primordial Black Holes (PBHs) \cite{Sasaki:2018dmp,Carr:2020xqk,Barausse:2020rsu,Addazi:2021xuf,Aalbers:2022dzr,Braglia:2022phb}, of cosmic strings and of other defects or ECOs that can be efficiently addressed within our analytic approach, very much as for the spectral index of density perturbations \cite{Sasaki:1995aw}. 
Once again gauge theory and (quantum) gravity can be beautifully connected with String Theory bridges.

\section*{Acknowledgements}

We would like to especially thank A. Balbi, A.~Cipriani, G.~Di~Russo, M. Migliaccio and N. Vittorio  for useful discussions and comments and M.~Sasaki for valuable suggestions. We also acknowledge fruitful scientific exchange with A.~Linde, P.~Pani, K.~Skenderis, J.~Silk and all the participants in the Lemaitre conference at Specola Vaticana in June and Pope Francis for very inspiring addresses. 
M.~B. and J.~F.~M. thank the MIUR PRIN contract 2020KR4KN2 ``String Theory as a bridge
between Gauge Theories and Quantum Gravity'' and the INFN project ST\&FI `` String Theory and
Fundamental Interactions'' for partial support.

\appendix
\section{Explicit examples}
\label{app:appendix}

In this section we display the potentials $Q(a)$ for all multi-component universes in tables \ref{multiscalar} and   \ref{multitens}.

\subsection*{Scalar perturbations}

The case with $N=3$ singularities can be solved in terms  of hypergeometric functions, while for $N=4$ one finds a Heun Equation. For simplicity we write
\be
\kappa\,=\,{\hat{\kappa}H_0^2 \over 3 M_{\rm Pl}^2} \ , \qquad \textrm{and} \qquad \hat{k} \,=\,\frac{k}{H_0}\ .
\ee
The $Q$-function appearing in equation \eqref{eqcan} (canonical form), for each case is given by
{\small
\bea
Q_{\Lambda \kappa } &=& -\frac{9 \Omega _{\Lambda } \left(3 a^2 \Omega _{\Lambda }-2 \hat{\kappa} \right)}{4 \left( \hat{\kappa} -3 a^2 \Omega _{\Lambda }\right){}^2} \nn\\
Q_{\Lambda m } &=&   \frac{11 \Omega _m^2+4 z \Omega _{\Lambda } \Omega _m+20 z^2 \Omega _{\Lambda }^2}{144 z^2 \left(\Omega _m+z \Omega _{\Lambda }\right){}^2} \qquad , \qquad z=a^3 \nn\\
Q_{\kappa  m} &=& \frac{9 \Omega _m \left(8 a \hat{\kappa} -21 \Omega _m\right)}{16 a^2 \left(a \hat{\kappa} -3 \Omega _m\right){}^2} \nn  \\
Q_{\kappa  \gamma} &=&  \frac{a^4 \hat{\kappa}  \left(\hat{\kappa} -4 k^2\right)+6 a^2 \Omega _{\gamma } \left(5 \hat{\kappa} +2 k^2\right)-72 \Omega _{\gamma }^2}{4 \left(a^3 \hat{\kappa}
   -3 a \Omega _{\gamma }\right){}^2} \nn\\
Q_{\Lambda \gamma} &=&   \frac{-15 \Omega _{\gamma }^2+z^3 \Omega _{\Lambda } \left(4 k^2-15 z \Omega _{\Lambda }\right)+2 z \Omega _{\gamma } \left(2 k^2-33 z
   \Omega _{\Lambda }\right)}{48 z^2 \left(\Omega _{\gamma }+z^2 \Omega _{\Lambda }\right){}^2}  \qquad , \qquad z=a^2\nn\\
Q_{m  \gamma} &=&  \frac{ P_{m\gamma}(a) }{48 a^2 \left(a \Omega _m+\Omega _{\gamma }\right){}^2 \left(3 a \Omega _m+4 \Omega _{\gamma }\right){}^2} \nn
   \eea
   \bea
  Q_{\Lambda\kappa m} &=& -\frac{9 \left(\Omega _m \left(60 a^3 \Omega _{\Lambda }-8 a \hat{\kappa} \right)+4 a^4 \Omega _{\Lambda } \left(3 a^2 \Omega _{\Lambda }-2 \hat{\kappa}
   \right)+21 \Omega _m^2\right)}{16 a^2 \left(-3 a^3 \Omega _{\Lambda }+a \hat{\kappa} -3 \Omega _m\right){}^2} \nn\\
 Q_{\Lambda\kappa \gamma} &=& \frac{-45 \Omega _{\gamma }^2+z^2 \left(4 \hat{\kappa}  \left(\hat{\kappa} -k^2\right)+12 z \left(\hat{\kappa} +k^2\right) \Omega _{\Lambda }-45 z^2 \Omega
   _{\Lambda }^2\right)+6 z \Omega _{\gamma } \left(2 \left(\hat{\kappa} +k^2\right)-33 z \Omega _{\Lambda }\right)}{16 z^2 \left(3 \Omega
   _{\gamma }+z \left(3 z \Omega _{\Lambda }-\hat{\kappa} \right)\right){}^2} ,~z=a^2 \nn\\
   Q_{\kappa m  \gamma} &=&  \frac{P_{ \hat{\kappa} m\gamma} (a)}{16 a^2 \left(3 a \Omega
   _m+4 \Omega _{\gamma }\right){}^2 \left(a^2 \hat{\kappa} -3 a \Omega _m-3 \Omega _{\gamma }\right){}^2}\nn\\
   Q_{\Lambda m  \gamma} &=& -\frac{ P_{ \Lambda m\gamma} (a) }{48 a^2 \left(3 a \Omega
   _m+4 \Omega _{\gamma }\right){}^2 \left(a^4 \Omega _{\Lambda }+a \Omega _m+\Omega _{\gamma }\right){}^2}   \nn\\
    Q_{\Lambda \kappa m  \gamma} &=&  -\frac{P_{ \Lambda m\gamma \hat{\kappa}} (a) }{16 a^2 \left(3 a \Omega _m+4 \Omega _{\gamma }\right){}^2 \left(3 a^4 \Omega _{\Lambda }-a^2 \hat{\kappa} +3 a \Omega _m+3 \Omega _{\gamma
   }\right){}^2}
   \eea 
} 
with
\bea
P_{m\gamma} (a) &=&  -567 a^4 \Omega _m^4-2772 a^3 \Omega _{\gamma } \Omega _m^3+256 \Omega _{\gamma }^3 \left(a^2 k^2-6 \Omega _{\gamma }\right)+12 a^2
   \Omega _{\gamma } \Omega _m^2 \left(16 a^2 k^2-455 \Omega _{\gamma }\right)\nn\\
   && +64 a \Omega _{\gamma }^2 \Omega _m \left(7 a^2 k^2-75 \Omega
   _{\gamma }\right) \nn\\
  P_{ \hat{\kappa} m\gamma} (a) &=&  24 a^5 \hat{\kappa}  \Omega _m \left(8 \Omega _{\gamma } \left(\hat{\kappa} -k^2\right)+27 \Omega _m^2\right)+a^4 \left(64 \hat{\kappa}  \Omega _{\gamma }^2
   \left(\hat{\kappa} -4 k^2\right)+72 \Omega _{\gamma } \left(29 \hat{\kappa} +8 k^2\right) \Omega _m^2-1701 \Omega _m^4\right) \nn\\
   &&+84 a^3 \Omega _{\gamma
   } \Omega _m \left(8 \Omega _{\gamma } \left(5 \hat{\kappa} +2 k^2\right)-99 \Omega _m^2\right)+12 a^2 \Omega _{\gamma }^2 \left(32 \Omega
   _{\gamma } \left(5 \hat{\kappa} +2 k^2\right)-1365 \Omega _m^2\right) \nn\\
   && -14400 a \Omega _{\gamma }^3 \Omega _m-4608 \Omega _{\gamma }^4 \nn\\
    P_{ \Lambda m\gamma} (a) &=&  324 a^{10} \Omega _{\Lambda }^2 \Omega _m^2+1152 a^9 \Omega _{\gamma } \Omega _{\Lambda }^2 \Omega _m+1536 a^8 \Omega _{\gamma }^2 \Omega
   _{\Lambda }^2+12 a^7 \Omega _{\Lambda } \Omega _m \left(135 \Omega _m^2-16 k^2 \Omega _{\gamma }\right) \nn\\
   &&+8 a^6 \Omega _{\gamma } \Omega
   _{\Lambda } \left(855 \Omega _m^2-32 k^2 \Omega _{\gamma }\right)+10560 a^5 \Omega _{\gamma }^2 \Omega _{\Lambda } \Omega _m+a^4 \left(5376
   \Omega _{\gamma }^3 \Omega _{\Lambda }-192 k^2 \Omega _{\gamma } \Omega _m^2+567 \Omega _m^4\right) \nn\\
   && +28 a^3 \Omega _{\gamma } \Omega _m
   \left(99 \Omega _m^2-16 k^2 \Omega _{\gamma }\right)+4 a^2 \Omega _{\gamma }^2 \left(1365 \Omega _m^2-64 k^2 \Omega _{\gamma }\right)+4800
   a \Omega _{\gamma }^3 \Omega _m+1536 \Omega _{\gamma }^4  \nn\\
P_{ \Lambda \kappa m\gamma} (a) &=&  972 a^{10} \Omega _{\Lambda }^2 \Omega _m^2+3456 a^9 \Omega _{\gamma } \Omega _{\Lambda }^2 \Omega _m-72 a^8 \Omega _{\Lambda } \left(9 \hat{\kappa}
    \Omega _m^2-64 \Omega _{\gamma }^2 \Omega _{\Lambda }\right) \nn\\
    && +36 a^7 \Omega _{\Lambda } \Omega _m \left(-40 \hat{\kappa}  \Omega _{\gamma }-16
   k^2 \Omega _{\gamma }+135 \Omega _m^2\right)+24 a^6 \Omega _{\gamma } \Omega _{\Lambda } \left(-80 \hat{\kappa}  \Omega _{\gamma }-32 k^2 \Omega
   _{\gamma }+855 \Omega _m^2\right) \nn\\
   && -24 a^5 \Omega _m \left(8 \hat{\kappa} ^2 \Omega _{\gamma }-1320 \Omega _{\gamma }^2 \Omega _{\Lambda }-8 \hat{\kappa}
    k^2 \Omega _{\gamma }+27 \hat{\kappa}  \Omega _m^2\right) \nn\\
    && +a^4 \left(-64 \hat{\kappa} ^2 \Omega _{\gamma }^2+16128 \Omega _{\gamma }^3 \Omega _{\Lambda
   }+256 \hat{\kappa}  k^2 \Omega _{\gamma }^2-576 k^2 \Omega _{\gamma } \Omega _m^2-2088 \hat{\kappa}  \Omega _{\gamma } \Omega _m^2+1701 \Omega
   _m^4\right) \nn\\
   && +84 a^3 \Omega _{\gamma } \Omega _m \left(-40 \hat{\kappa}  \Omega _{\gamma }-16 k^2 \Omega _{\gamma }+99 \Omega _m^2\right)+12 a^2
   \Omega _{\gamma }^2 \left(-160 \hat{\kappa}  \Omega _{\gamma }-64 k^2 \Omega _{\gamma }+1365 \Omega _m^2\right) \nn\\
   && +14400 a \Omega _{\gamma }^3
   \Omega _m+4608 \Omega _{\gamma }^4
\eea

    \subsubsection*{Tensor perturbations}
 
For universes made of two or three   $\L$CDM  components one finds the potentials
\bea
Q_{\textrm{T},\gamma m}(a) &=& \frac{8 \Omega _m \left(2 a^2 k^2-\Omega _{\gamma }\right)+16 a k^2 \Omega _{\gamma }-5 a \Omega _m^2}{16 a \left(a \Omega _m+\Omega
   _{\gamma }\right){}^2} \nn \\
 Q_{\textrm{T},\Lambda\gamma }(a) &=&  -\frac{-3 \Omega _{\gamma }^2-4 k^2 z^3 \Omega _{\Lambda }-4 k^2 z \Omega _{\gamma }+5 z^4 \Omega _{\Lambda }^2+14 z^2 \Omega _{\gamma }
   \Omega _{\Lambda }}{16 z^2 \left(\Omega _{\gamma }+z^2 \Omega _{\Lambda }\right){}^2}  \quad z=a^2\nn\\
   Q_{\textrm{T},\Lambda m}(a) &=&\frac{16 a \Omega _m \left(k^2-4 a^2 \Omega _{\Lambda }\right)+16 a^4 \Omega _{\Lambda } \left(k^2-2 a^2 \Omega _{\Lambda }\right)-5 \Omega
   _m^2}{16 a^2 \left(a^3 \Omega _{\Lambda }+\Omega _m\right){}^2}  \\
  Q_{\textrm{T},\Lambda m \gamma}(a) &=&  \frac{16 a \left(\Omega _{\gamma } \left(k^2{-}5 a^2 \Omega _{\Lambda }\right){+}a^4 \Omega _{\Lambda } \left(k^2{-}2 a^2 \Omega _{\Lambda
   }\right)\right){-}8 \Omega _m \left(8 a^4 \Omega _{\Lambda }{-}2 a^2 k^2{+}\Omega _{\gamma }\right){-}5 a \Omega _m^2}{16 a \left(a^4 \Omega
   _{\Lambda }{+}a \Omega _m{+}\Omega _{\gamma }\right){}^2}\nn
\eea 
  
 \section{Cosmic time vs scalar factor for various multi-component fluids}\label{Cosmic_t}
     
% \section{What's the (cosmic) time?}
%
% As we mentioned, cosmic time $t$ tends to complicate the analysis of both the background solution and the linear perturbations. Yet we should check that $a(t)$ is monotonous if we want it to be a 'good' time variable. The relation between $t$ and $a$ is determined by
% \be
% dt = {da \over b(a)} = {da \over a H(a)} =  {da \over a H_0 \sqrt{\sum\limits_{i}   \Omega_i  a^{-3(1+w_i)}  +\Omega_\kappa a^{-2}}}
% \ee
%In general the integral
%\be
%H_0 t = \int {da \over \sqrt{\Omega_\Lambda a^2 +\Omega_\kappa + \Omega_{\rm m} a^{-1} + \Omega_\gamma a^{-2}}} =  \int {a da \over \sqrt{\Omega_\Lambda a^4 +\Omega_\kappa a^2 + \Omega_{\rm m} a + \Omega_\gamma}}
%\ee
%cannot be expressed in terms of elementary functions except for single and two-component fluids.
In the following we display the relation between cosmic time $t$ and scale factor $a$ for various cases with both $\kappa =0$ and $\kappa\neq 0$.

{\bf $\kappa =0$}
\begin{itemize}
\item{$\Lambda$ $\gamma$: vacuum+radiation}
\be
H_0 t = \int {a da \over \sqrt{\Omega_\Lambda a^4 + \Omega_\gamma } } =
{1\over 2\sqrt{\Omega_\Lambda}} \log{
\sqrt{\Omega_\Lambda} a^2 + \sqrt{
\Omega_\Lambda a^4 +\Omega_\gamma} \over \sqrt{\Omega_\gamma}}
 \ee
 or
 \be
 a^2 = \sqrt{\Omega_\gamma \over \Omega_\Lambda}\sinh (2\sqrt{\Omega_\Lambda} H_0 t)
 \ee
 \item{{m} + $\Lambda$: matter+vacuum}
\be
H_0 t = \int { a^2 da \over \sqrt{\Omega_\Lambda a^6 + \Omega_{\rm m} a^3} } =
 {1\over 3\sqrt{\Omega_\Lambda}} \log
 {2\Omega_\Lambda a^3 +\Omega_{\rm m} + \sqrt{\Omega_\Lambda a^6 + \Omega_{\rm m}a^3} \over \Omega_{\rm m}}
 \ee
 or
 \be
 a^3 = {\Omega_{\rm m} \over \Omega_\Lambda }  \sinh^2\left({3 \over 2}\sqrt{\Omega_\Lambda} H_0 t\right)
 \ee
\item{$\gamma$ + m: radiation+matter}
\be
H_0 t = \int { a da \over \sqrt{\Omega_\gamma + \Omega_{\rm m} a} } = {2 \over 3 \Omega_{\rm m}^2} \left(2 \Omega_\gamma^{3/2} + (\Omega_{\rm m} a-2\Omega_\gamma) \sqrt{\Omega_{\rm m} a+\Omega_\gamma}\right)  \ee
\end{itemize}

{\bf $\kappa \neq 0$}
\begin{itemize}
\item{$\kappa$, $\Lambda$, $\gamma$: curvature+vacuum+radiation}
\be
H_0 t = \int {a da \over \sqrt{\Omega_\Lambda a^4 + \Omega_\kappa a^2 + \Omega_\gamma } } =
 {1 \over 2\sqrt{\Omega_\Lambda}} \log { 2\Omega_\Lambda a^2 + \Omega_\kappa + 2\sqrt{\Omega_\Lambda}\sqrt{\Omega_\Lambda a^4 + \Omega_\kappa a^2 + \Omega_\gamma } \over  \Omega_\kappa + 2\sqrt{\Omega_\Lambda \Omega_\gamma }}
\ee
\item{$\kappa$, m, $\Lambda$: curvature+matter+vacuum}
\be
H_0 t = \int { a da \over \sqrt{\Omega_\Lambda a^4 + \Omega_\kappa a^2+ \Omega_{\rm m} a} } =  {\rm elliptic \, NON \, elementary}
 \ee
\item{$\kappa$, $\gamma$, m: curvature+radiation+matter}
\be
H_0 t = \int { a da \over \sqrt{\Omega_\gamma + \Omega_{\rm m} a + \Omega_\kappa a^2} } =
 \ee
 $$
 = {1 \over \Omega_\kappa} \left(
 \sqrt{\Omega_\gamma + \Omega_{\rm m} a + \Omega_\kappa a^2} - \sqrt{\Omega_\gamma} - {\Omega_{\rm m} \over 2 \sqrt{\Omega_\kappa} }
 \log{ 2\Omega_\kappa a + \Omega_{\rm m} + 2 \sqrt{\Omega_\kappa}\sqrt{\Omega_\gamma + \Omega_{\rm m} a + \Omega_\kappa a^2} \over \Omega_{\rm m} + 2 \sqrt{\Omega_\kappa\Omega_\gamma}} \right)
 $$
\end{itemize}

%In the analysis of linear perturbations we will only consider adiabatic modes and neglect entropy or iso-curvature modes. It is known\footnote{We thank Misao Saski for reminding us about this issue} \cite{Kodama:1984ziu, Mukhanov:1990me, Vittorio:2017foh} that adiabatic
%modes are necessarily coupled to entropy (or isocurvature) modes in the case of a multi-component fluid. In particular the coupling with these modes introduces non-homogeneous source terms in the equations for the adiabatic modes that are particularly relevant during the transient from one fluid phase to another. As a result the exact knowledge of the solutions of the homogeneous equations for purely adiabatic modes may not describe the actual behavior of an adiabatic perturbation, such as the scalar $\Phi$ or the tensor $h$. In order to estimate
%this effect, one should know the strength of the coupling to entropy and iso-curvature modes, which would require determining the first non-linear corrections to the linear homogeneous equations and may be rather model dependent. Yet, treating the coupling as a non-linear correction allows to resort to the Green function method that would benefit a lot from our knowledge of the analytic solutions to the homogeneous equations. We hope to come back to this point.

\bibliographystyle{JHEP}
\bibliography{Ref}

\end{document}